\documentclass[aps,twocolumn,amsmath,amssymb,prb,superscriptaddress,longbibliography]{revtex4-1} 
\usepackage{graphicx}
\usepackage{txfonts}
\usepackage{mathptmx}
\usepackage{braket}
\usepackage{color}
\usepackage[breaklinks,colorlinks,bookmarks=false,citecolor=blue,linkcolor=blue,urlcolor=blue]{hyperref}
\pdfmapfile{+txfonts.map}

\begin{document}
\title{
Electronic structure and quantum transport in twisted bilayer graphene with resonant scatterers \\
}

\author{Omid Faizy Namarvar } \email{omid.faizy@xlim.fr}
%https://orcid.org/0000-0001-7605-3182
\affiliation{%Institut N\'eel, CNRS - Universit\'e Joseph Fourier, F-38042 Grenoble, France
%Institut N\' eel, CNRS/UJF, 25 rue des Martyrs BP166, 38042 Grenoble Cedex 9, France.
Univ. Grenoble Alpes, CNRS, Institut N\'eel, 38000 Grenoble France}
\affiliation{
%Centro S3, CNR - Istituto Nanoscienze, I-41125 Modena, Italy
XLIM, Univ. Limoges, CNRS UMR 7252, 87000 Limoges, France
}
\author{Ahmed Missaoui }  \email{ahmed.missaoui@fst.utm.tn}
\affiliation{Laboratoire de Spectroscopie Atomique Mol\'eculaire et Applications, 
D\'epartement de Physique, Facult\'e des Sciences de Tunis, Universit\'e Tunis El Manar,  Campus Universitaire 1060 Tunis, Tunisia}
\affiliation{CY Cergy Paris Universit\'e, CNRS, Laboratoire de Physique th\'eorique et Mod\'elisation,
95302 Cergy-Pontoise, France}
\author{Laurence Magaud }  \email{laurence.magaud@neel.cnrs.fr}
\author{Didier Mayou }  \email{didier.mayou@neel.cnrs.fr}
\affiliation{%Institut N\'eel, CNRS - Universit\'e Joseph Fourier, F-38042 Grenoble, France
%Institut N\' eel, CNRS/UJF, 25 rue des Martyrs BP166, 38042 Grenoble Cedex 9, France.
Univ. Grenoble Alpes, CNRS, Institut N\'eel, 38000 Grenoble France}
\author{Guy Trambly de Laissardi\`ere}  \email{guy.trambly@cyu.fr}
\affiliation{%Institut N\'eel, CNRS - Universit\'e Joseph Fourier, F-38042 Grenoble, France
%Institut N\' eel, CNRS/UJF, 25 rue des Martyrs BP166, 38042 Grenoble Cedex 9, France.
CY Cergy Paris Universit\'e, CNRS, Laboratoire de Physique th\'eorique et Mod\'elisation,
95302 Cergy-Pontoise, France}

\
\date{\today{}}

\begin{abstract}
Staking layered materials revealed to be a very powerful method to tailor their electronic properties. It has indeed been theoretically and experimentally shown that twisted bilayers of graphene (tBLG) with a rotation angle $\theta$, forming Moir\'e pattern, confine electrons in a tunable way as a function of $\theta$.
Here, we study electronic structure and transport in tBLG  using  tight-binding numerical calculations in commensurate twisted bilayer structures and a pertubative continuous theory, which is valid for not too small angles ($\theta > \sim 2^\circ $). These two approaches allow to understand the effect of $\theta$ on the local density of states, the electron lifetime due to disorder, the dc-conductivity and the conductivity quantum correction due to multiple scattering effects. 
We distinguish the cases where disorder is equally distributed over the two layers or only over one layer. 
When only one layer is disordered, diffusion properties depend strongly on $\theta$, showing thus the effect of Moir\'e electronic localisation at intermediate angles $ \theta$, $\sim 2^\circ < \theta <  \sim 20^\circ$. 
%\textcolor{red}{??? Une phrase sur les applications potentielles ????  }

%We calculate the self-energy of states of the upper plane due to their coupling with states of the lower plane . We discuss consequences for velocity renormalization %and for electron-lifetime due to disorder in one plane. We also calculate and discuss the spatial modulations of density of states.
%We compare our analytical results to fully numerical calculations, showing good agreement between the two approaches.

\end{abstract}

%\pacs{73.22.Pr; 73.21.Ac}

\maketitle

%\tableofcontents

\section{Introduction}

%\textcolor{red}{ }

Staking layered materials is a very powerful method to tailor their electronic properties.\cite{Geim13} The properties not only depend on the choice of materials to be stacked but also on the details of the relative arrangement of the layers. It has thus been theoretically\cite{LopesdosSantos07,Trambly10,SuarezMorell10,Bistritzer10,Bistritzer11,Trambly12} and experimentally\cite{Li09,Luican11,Brihuega12,Huder18} shown that twisted bilayer graphene (tBLG), forming Moir\'e pattern, confine conduction electrons in a tunable way as a function of the angle of rotation of one layer with respect to the other. Recently, it has been experimentally proven that this electronic localization by geometry can induce strong electronic correlations\cite{Cao18_correlated} and a superconducting state\cite{Cao18_unconventional} for certain angles called magic angles.\cite{Bistritzer11} 
Despite numerous studies of the electronic structure of these systems,\cite{LopesdosSantos07,Latil07,Trambly10,SuarezMorell10,Bistritzer10,Bistritzer11,
Trambly12,Li09,Luican11,Brihuega12,Huder18,Brihuega12,LopesdosSantos12,Trambly16,Chari16,Le18,
Chung18,Wu18,RibeiroPalau18,Andelkovic18,Jeon18,Rickhaus18,Wu18,Chung18,Le18,Cao18_correlated,
Cao18_unconventional,Do19,Sharma20,Hidalgo19} 
the consequences of the electronic localization by a Moir\'e on electrical transport properties are still poorly known. 
In particular the effects of local defects such as adsorbated atoms or adsorbated molecules, which are known to tune strongly electronic properties in graphene based 2D materials.\cite{Katoch18,Hidalgo19,PINTO20}

Graphene can be formed in multilayers on SiC \cite{Ohta06,Coletti10, Brihuega08,Hass06,Hass08b,Emtsev08,Hass08,Sprinkle09,Hicks11} but also on metal surfaces such as Ni \cite{Luican11} and in exfoliated flakes,\cite{Li09}  where hopping terms between successive layers play a crucial role. While on the Si face of SiC, multilayers have an AB Bernal stacking and do not show graphene properties,\cite{Latil06,Ohta06,Brihuega08,Varchon08,Coletti10,Zhang10,McCann13,Rozhkov16} on the C-face multilayers are twisted multilayers of graphene  with various angles of rotation between two successive layers. 
For large twist angle $\theta$ between two layers, multilayers 
show graphene-like properties even when they involve a large number of graphene layers. Indeed, as shown by ARPES,\cite{Emtsev08,Hass08,Sprinkle09,Hicks11} STM,\cite{Miller09} transport\cite{Berger06} and optical transitions,\cite{Sadowski06} their properties are characteristic of a linear graphene-like dispersion. 
Therefore, in tBLG interlayer hopping terms does not systematically destroy graphene like properties, but it can lead to the emergence of very peculiar and new behaviors induced by the Moir\'{e} patterns that is accentuated for $\theta$ smaller than $\sim 20\rm ^\circ$.
Theoretical studies have predicted \cite{LopesdosSantos07,Trambly10,SuarezMorell10,Bistritzer10,Bistritzer11,Trambly12,LopesdosSantos12} the existence of three domains: 
$(1)$ for large rotation angles ($\theta > 20 ^\circ $) the layers are decoupled and behave as a collection of isolated graphene layers. 
$(2)$ For intermediate angles $ \sim 2^\circ < \theta< 20^\circ $ the dispersion, around Fermi energy $E_F$, remains linear but the velocity is renormalized. 
Consequently, the energies of the two van Hove singularities $E_-$ and $E_+$ are shifted to Dirac energy $E_D$ when $\theta$ decreases, as it has been shown experimentally. \cite{Luican11,Brihuega12,Ohta12,Cherkez15} 
$(3)$ For the lowest $\theta$, $\theta< \sim 2^\circ$,  almost flat bands appear and result in electronic localization in AA stacking regions: states of similar energies, belonging to the Dirac cones of the two layers interact strongly.
In this regime, the velocity of states at Dirac point goes to almost zero for specific angle so-called magic angles.\cite{Trambly10,Bistritzer11,Trambly12}  
Recently, the signature of the electron localization in the AA regions at long time evolution has been confirmed numerically for small $\theta$.\cite{Do19}

In this paper, we study the consequence of the tunable effective coupling between layers by angle  $\theta$ with intermediate values, $\sim  2^\circ < \theta< \sim 20^\circ $, on local density of states (LDOS) and transport  properties. 
We combine tight-binding (TB) numerical calculations for commensurate tBLG and a perturbative continuous theory (see Appendix) that gives us deeper insight on $\theta$ effect.  
Note that our TB calculation include all matrix element couplings; whereas the continuous theory,
like the one previously developed,
\cite{LopesdosSantos07,LopesdosSantos12} neglect the coupling of electrons in different valleys. 
To analyze transport properties numerically in bulk 2D systems, 
we consider local defects,\cite{Lazar13,Brihuega18} such as adsorbates or vacancies, that are resonant scatterers. Local defects tend
to scatter electrons in an isotropic way for each valley and lead also to strong intervalley scattering. 
The adsorbate is simulated by a simple vacancy in the layer of p$_z$ orbital as usually
done.\cite{Castro09_RevModPhys,Trambly13,Missaoui17} 
Indeed the covalent bonding between the adsorbate and the carbon atom of graphene to which it is linked, eliminates the p$_z$ orbital from the relevant
energy window.
We consider here that the up and down spins are degenerate, i.e. we deal with a paramagnetic state. Indeed
the existence and the effect of a magnetic state for various adsorbates or vacancies is still debated.\cite{Nair12,Scopel16} 
In the case of a magnetic state the up and down spins give two different contributions to the conductivity but the individual contribution of 
each spin can be analyzed from the results discussed here.
We consider the case  $(i)$ where defects are located in the two layers, with respect to the case $(ii)$ where defects are located on one layer (layer 2) only. 

In Sec. \ref{Sec_DOS}, tight-binding (TB)  local Density of states (LDOS) in pristine tBLG and the effect of disorder on total DOS (TDOS) are analyzed with respect to our analytical model for commensurate tBLG. 
The spatial modulation of the DOS shows an increase of the DOS in AA region of the Moir\'{e}. This is a precursor of the localization in the AA region for very small angles less than $ \sim 2^\circ$.\cite{Trambly10,Trambly12}
The
electrical dc-conductivity at high temperature (microscopic conductivity) is studied in  Sec. \ref{Sec_transpHT}, and 
quantum corrections of conductivity (low temperature limit) are presented in Sec \ref{Sec_transpLT}. 
The method to compute dc-conductivity is given in the appendix \ref{Sec_Kubo}.
Numerical resuts of the paper are analyzed using the analytical continous model presented in appendix \ref{Sec_TB} and \ref{Sec_IC}. 
This pertubative theory recovers known results for the velocity renormalization,\cite{LopesdosSantos07,LopesdosSantos12} 
but provides new analytical results concerning 
LDOS and state lifetime  versus $\theta$ values. 

The method to built commensurate tBLG is well known and explained in many articles. Here we use the notations used in our previous papers \cite{Trambly10,Trambly12,Trambly16} where each tBLG is built from two index $n$ and $m$ (table \ref{Tab_bilayer}). For $|m-n|=1$ the cell of the bilayer contains one Moir\'e cell, whereas for  $|m-n|>1$ the cell of the bilayer contains several Moir\'e cells. 

\begin{table}
\caption{\label{Tab_bilayer} 
Studied $(n,m)$ bilayer structures. $N$ is the number of atoms, $\theta$ the rotation angle.
}
\begin{tabular}{lrr}
\hline \hline
{($n,m$)}~~~  &  ~~~~~~~$\theta$ [deg.]   &  ~~~~~~~~~~~~~~~~$N$    \\ \hline
(12,13)    &  2.656     &  1876       \\
(10,11)   &  3.150  &  1324      \\
(8,9)    &  3.890     &  868        \\
(6,7)    &  5.086     &  508       \\
(5,6)    &  6.009     &  364      \\
(4,5)    &  7.341     &   244     \\
(7,9)    &  8.256  &   772  \\
%(10,13)  &  8.613  &   1596   \\
(10,13)  &  8.613  &   532  \\
(3,4)    &   9.430    &  148        \\
%(8,11)    &   10.417     &    1092      \\
(8,11)    &   10.417     &    364      \\
(2,3)    &   13.174     &     76     \\
(5,9)    &   18.734     &    604      \\
(1,3)  &   32.204  &  52      \\
(1,4)  &  38.213  & 84  \\
\hline \hline
\end{tabular}
%\end{ruledtabular}
\end{table}

\section{Density of States}
\label{Sec_DOS}

\subsection{Without defect}

\begin{figure}

\includegraphics[width=8.5cm]{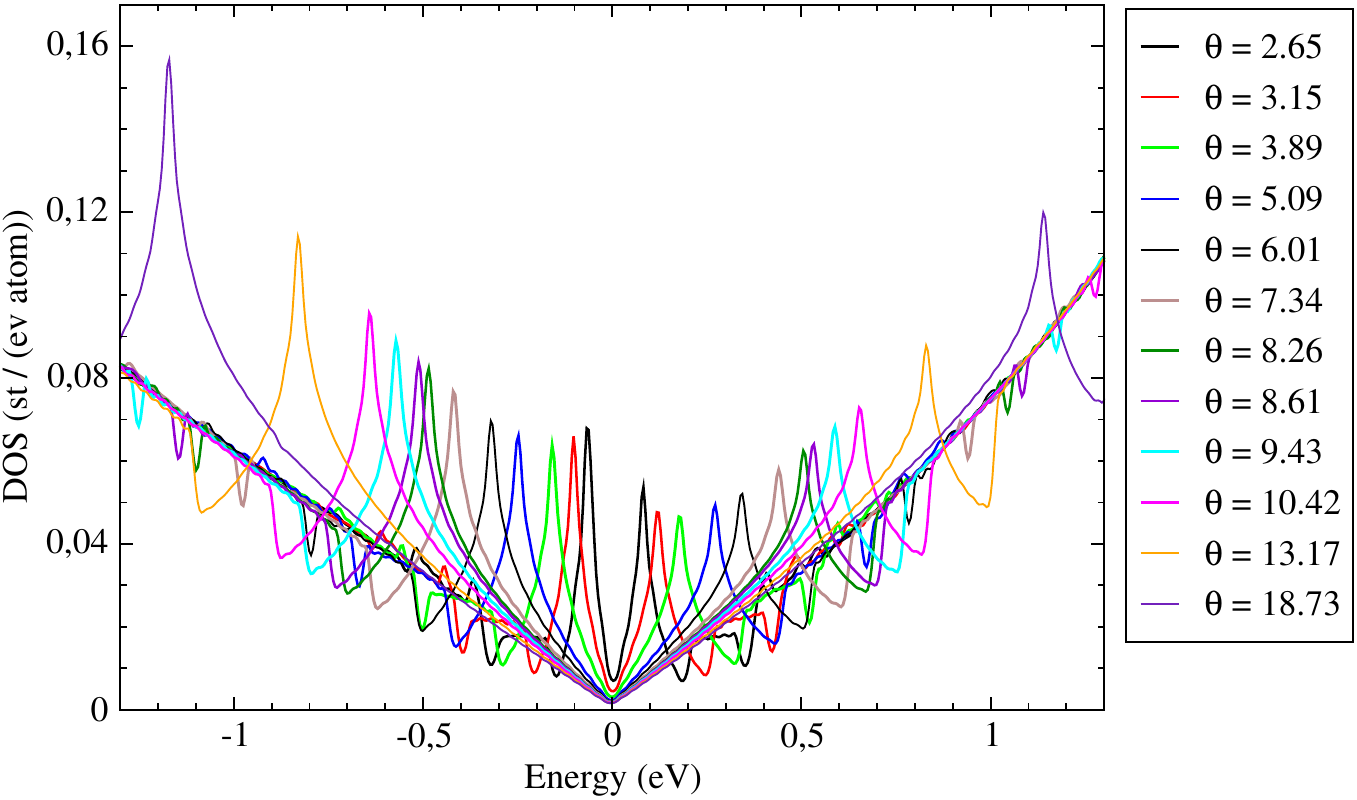} 

\caption{ \label{Fig_LDOS_AA}
(color online)
Local density of states (LDOS) at the center of a AA zone in pristine tBLG listed table \ref{Tab_bilayer} for tBLG with different rotation angle $\theta$ [Deg.].
Some LDOS are taken from Ref. \onlinecite{Trambly12}. $E_D=0$.
}
\end{figure}

We first analyze the local density of states (LDOS) in pristine twisted bilayer graphene (tBLG) computed with the TB Hamiltonian detailed in the Refs \onlinecite{Trambly12} and the appendix.  
It is now well known theoretically\cite{Trambly12,Brihuega12,LopesdosSantos12,Trambly16}  and experimentally\cite{Brihuega12,Cherkez15} that the energies $E_-$ and $E_+$ of Van Hove singularities vary linearly with the angle $\theta$ for $\theta > \sim 2 ^\circ$. 
This is clearly seen in the LDOS on p$_z$ orbital of atom located at the center of AA area of the Moir\'{e} (Fig. \ref{Fig_LDOS_AA}).
Since our TB Hamiltonian includes coupling beyond the first neighboring atoms, the electron / hole symmetry is slightly broken and $E_-$ is not strictly equal to $-E_+$. 

The LDOS in one layer of the bilayer as a function of position $\vec{r}$ in the Moir\'{e}  structure is 
\begin{equation}
\rho(E, \vec{r} )=\braket{\vec{r} |\delta (E-H) | \vec{r} }.
\end{equation}
To compare LDOS in bilayer with LDOS in monolayer we compute the relative variation of the LDOS due to interlayer hopping terms ${\Delta \rho(E,\vec{r})}/{\rho_m(E)}$, with $\Delta \rho(E,\vec{r}) = \rho(E,\vec{r}) - \rho_m(E)$, where $\rho_m(E)$ is the LDOS in monolayer that does not depend on the position $\vec r$. 

\begin{figure}[h]
\centering
%\vspace{2mm}
%\scalebox{0.21}{\includegraphics{DOS1.png}}
%\rotatebox{-90}{\resizebox{0.38\textwidth}{!}{\includegraphics{Fig_ldos_6-7_12-13.pdf}
%}}
\includegraphics[width=8.5cm]{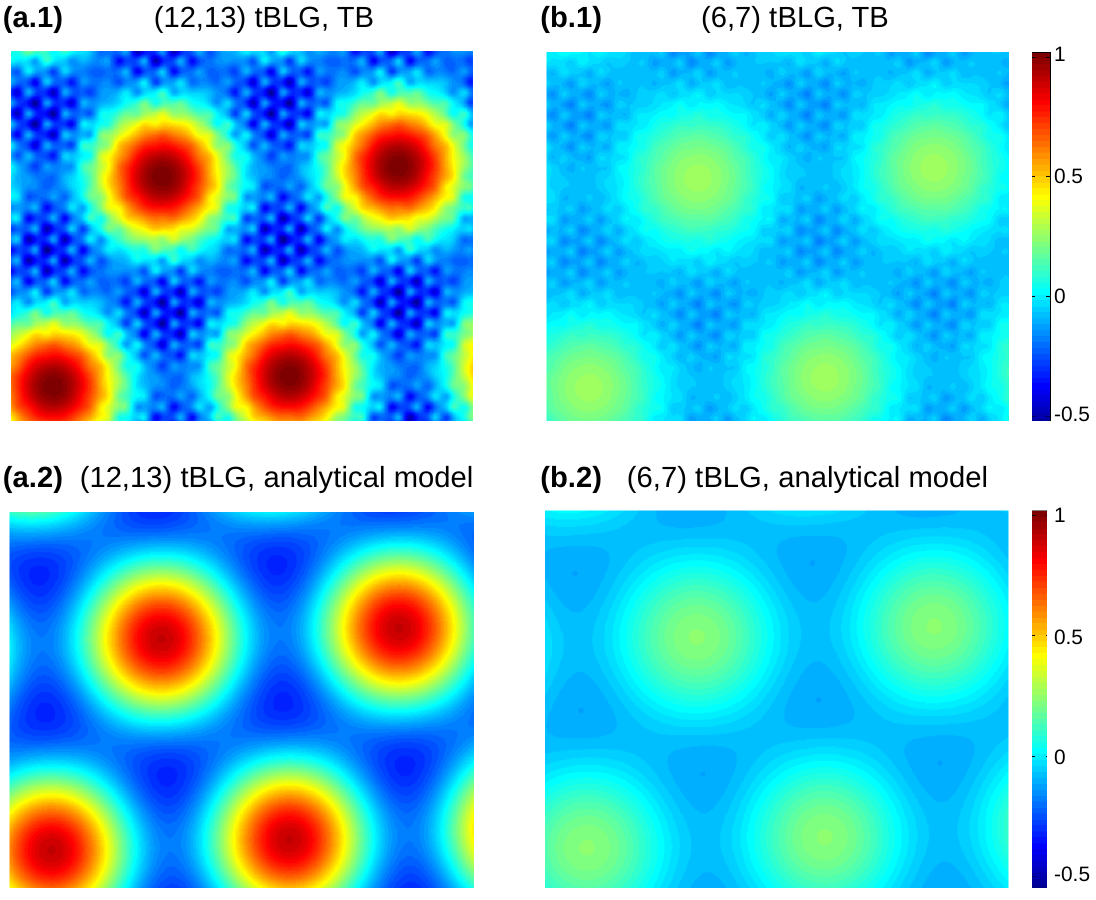}
\caption{\label{Fig_ldos6-7_12-13}
(Color online) Relative variation of the LDOS on top layer at energy $E=E_D+0.05$\,eV, close to the Dirac energy $E_D$, in (a) (12,13) tBLG and (b) (6,7) tBLG: 
(a.1) (b.1) TB results and (a.2) (b.2) analytic results from equation (\ref{eqLDOS}).
To be compared with analytic results the TB plots are made by a continuous extrapolation of LDOS on atomic orbitals.  
The same arbitrary unit are used for all the 4 LDOS. $E_D=0$.
}
\label{fig:rdos}
\end{figure}

The LDOS on each carbon atoms of Moir\'{e} has been calculated using TB, 
so that density map $\rho(E, \vec{r} )$ where $\vec r$ are the positions of Carbon atoms can be drawn for an energy $E$.  
Figures \ref{Fig_ldos6-7_12-13}(a.1) and \ref{Fig_ldos6-7_12-13}(b.1) show  relative TB LDOS in (12,13) and (6,7) bilayers at the energy $E = E_D + 0.05$\,eV. The strong increase of LDOS in AA areas with respect to the AB zone are clearly seen. As expected this difference between LDOS in AA area and AB area decreases as $ \theta$ increases. 
Moreover our numerical TB calculation recovers the difference in the LDOS of the two inequivalent atoms in AB area. Indeed in AB area, as in AB Bernal stacking, C atoms lying above a C atom of the other layer have a lower LDOS than the LDOS of C atom not lying above a C atom of the other layer.
That leads to a triangular contrast\cite{Tomanek87} in the density map that has been observed in STM images in AB Bernal bilayer.

According to the perturbative analytical model presented in Appendix (Sec. \ref{Sec_SpatialDOS}),  the relative variation of the LDOS is independant of $E$ for small $E$ and it can be estimated by the simple formula, 
\begin{equation}
\frac{\Delta \rho(E,\vec{r} )}{\rho_m(E)} \simeq  
\left( \frac{\theta_{1}}{\theta} \right)^2  \sum\limits_{j=1}^{6}  \, \cos(\vec{G_j} \cdot \vec{r}),
\label{eqLDOS}
\end{equation}
where $\vec{G_j}$ are  6 equivalent vectors of the reciprocal space of the Moir\'{e} lattice. The constant  $\theta_{1}$ is given by, 
\begin{equation}
\theta_{1}=\frac{\sqrt{2}t}{(\hbar v K_D )},
\label{theta1}
\end{equation}
where $v$ is the monolayer velocity and $K_D$ is the modulus of the wave-vector in Dirac point of graphene.
Using the interlayer coupling value $t \simeq 0.12$\,eV (Appendix Sec. \ref{Sec_Hamiltonian_R}), one finds that  the value of $\theta_{1}$ is close to $\theta_{1} \simeq 1^\circ$.
Equation (\ref{eqLDOS}) does not depend on the type of atom (atom A or atom B) it oscillates with $\vec{G}_j$ as expected. 
As it is clear the maximum value is obtained for $\vec{r}=0$ which is at the center of AA area, and relative variation of the LDOS varies as $\theta^{-2}$. 
%and $  \sum\limits_{j=1}^{6}  \,   \cos(\vec{G_j} \cdot \vec{R})  $ is 6. For example for $\theta =  3 ^\circ$ the maximum value of relation variation of DOS is $\frac{6}{9}$ in the AA region which is strong. 
%We note also that the relative variation rapidly decreases when $\theta$ increases.
As shown in Fig. \ref{Fig_ldos6-7_12-13} the overall agreement between TB numerical calculation and TB analytical model is very good. We just note a small triangular contrast in AB zone which is not reproduced by the analytical model 
(see Appendix Sec. \ref{Sec_SpatialDOS} for a discussion). We observe in particular a reinforcement of the DOS in the AA region and a lowering in the AB regions. This behavior is a precursor of the electronic localization in AA region which is observed in the very low angle limit $\theta < \sim 2^\circ$.\cite{Trambly10,Trambly12,Huder18} 

\subsection{With resonant adsorbates}

\begin{figure}

\includegraphics[width=4.2cm]{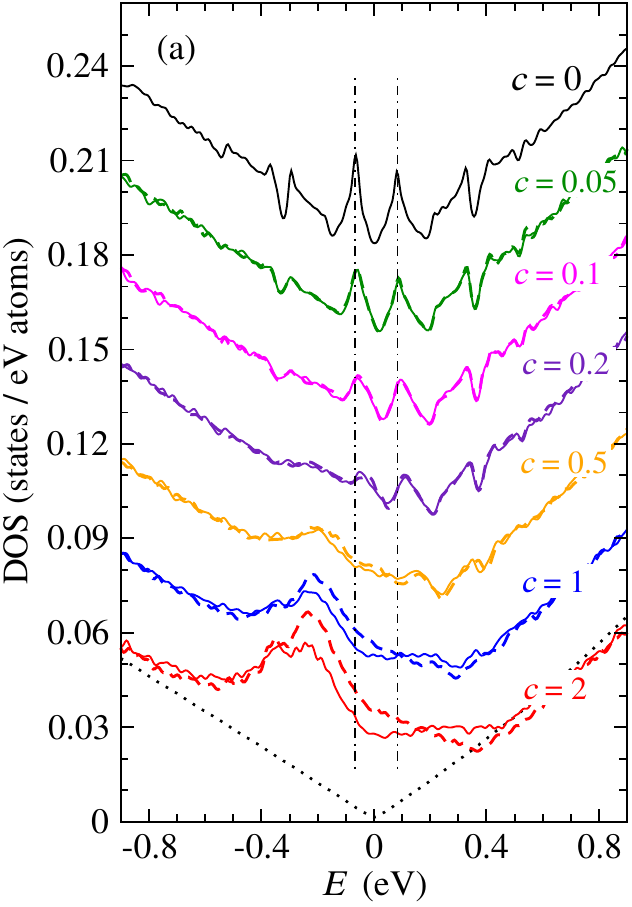} 
~ \includegraphics[width=4.2cm]{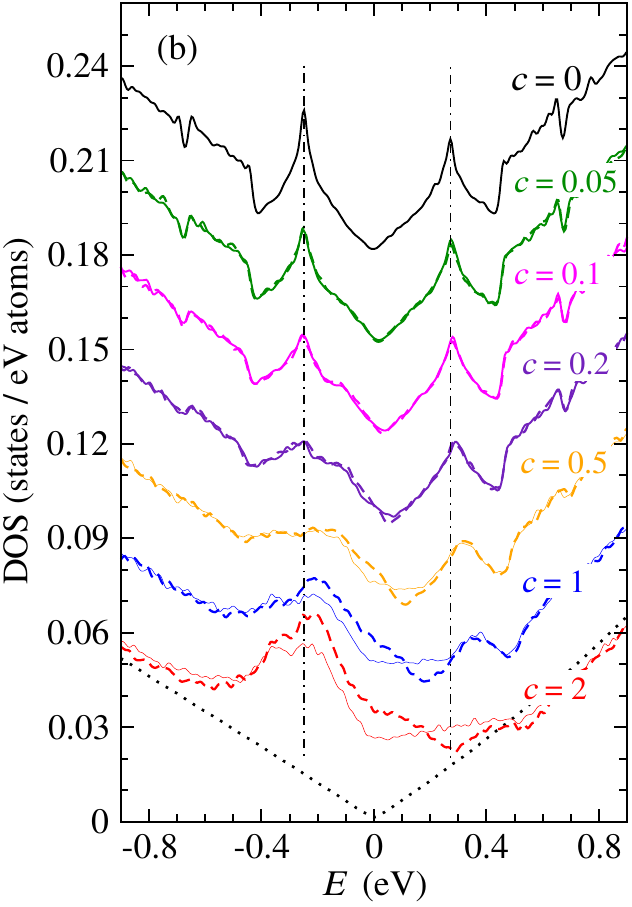} 

\caption{ \label{Fig_tDOS_Va}
(color online)
Total DOS in (a) (12,13) tBLG and (b) (6,7) tBLG, for various concentrations $c$ (\%) of vacancies with respect to the total 
number of atom in tBLG: 
(Dashed line) with vacancies in both layers and (full line) with vacancies in layer 2. 
(Dotted line) is the DOS in pristine monolayer graphene (MLG). $E_D=0$.
}
\end{figure}

\begin{figure}

\includegraphics[width=4.1cm]{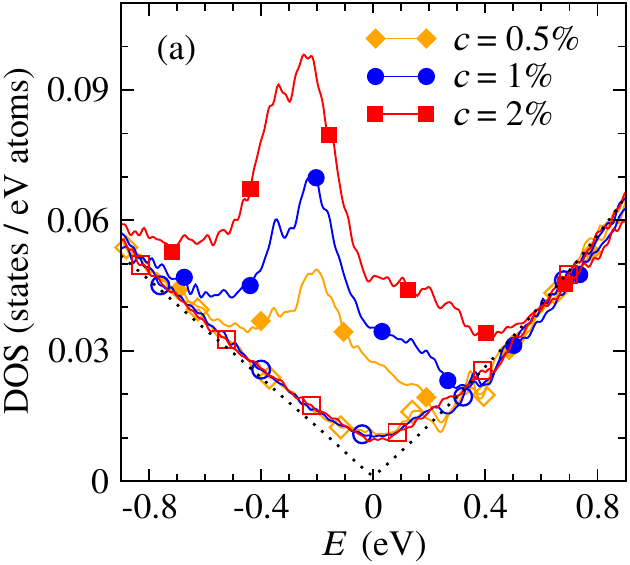} 
~ \includegraphics[width=4.1cm]{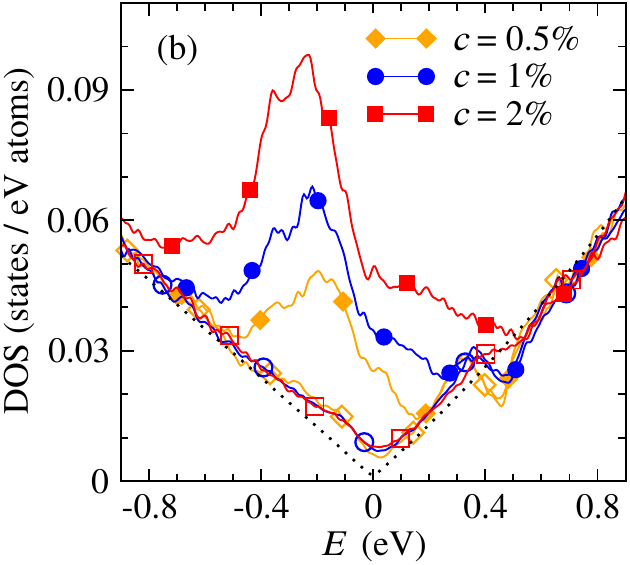} 

\caption{ \label{Fig_tDOS_P_Va}
(color online)
Average LDOS in each layer: (a) (12,13) tBLG and (b) (6,7) tBLG,  for various concentrations $c$ (\%) of vacancies in layer 2. 
(empty symbol) average LDOS in layer 1, (full symbol) average LDOS in layer 2.
(Dotted line) is the DOS in pristine monolayer graphene (MLG).
$c$ is the concentration of vacancies with respect to the total number of atom in tBLG. $E_D=0$.
}
\end{figure}

To study the effect of static defects on the electronic confinement by the Moir\'e we include atomic vacancies (vacant atoms) that simulate resonnant adsorbates atoms or molecules.\cite{Pereira06,Pereira08a,Wehling10,Trambly13,Trambly14,Fan14,Missaoui17,Missaoui18} 
For each vacancies concentrations $c$ with respect to the total number of Carbon atoms in tBLG, we consider two cases: 
\begin{itemize}
\item[$(i)$] vacancies are randomly distributed in both layers, 
\item[$(ii)$] vacancies are randomly distributed in layer 2 only. 
\end{itemize}

Total DOSs (tDOSs) in (12,13) tBLG and (6,7) tBLG are drawn Fig. \ref{Fig_tDOS_Va} for different concentrations of vacancies in cases $(i)$ and $(ii)$.
For small $c$ values, the Van Hove singularities are still clearly seen but they are enlarged by disorder.
This shows that static disorder destroys the confinement by Moir\'e in AA areas. 
For $c > \sim 0.5$\,\%, peaks of the Van Hove singularities are destroyed by vacancy states.
With TB Hamiltonian including only first neighbor hopping terms, the vacancy states are midgap states at Dirac energy.\cite{Pereira06,Pereira08a} 
But, as in monolayer graphene \cite{Trambly14} and Bernal bilayer graphene,\cite{Missaoui17}
taking into account the TB hoppings beyond first neighbor enlarges the midgap states and shifts it to negative energies, typically around $-0.2$\,eV. 
As shown in Fig. \ref{Fig_tDOS_P_Va}, when vacancies are located in layer 2 only (case $(ii)$), the vacancy states only appear  on LDOS
p$_z$ orbitals of layer 2. 
Note that average DOS in layer 1 is slightly modified by the vacancies located in layer 2 (Fig. \ref{Fig_tDOS_P_Va}). This effect seems similar to modification due to nonresonant scatterers.\cite{Trambly13}  
Figs. \ref{Fig_tDOS_Va} and \ref{Fig_tDOS_P_Va} show that, as far as the DOS is concerned and for rather large concentration of vacancies ($c>0.5$\,\%), the rotated angle $\theta$ does not change the effect of vacancies.  
As we will see in next section, the effect of $\theta$ is more pronounced on wave-packet quantum diffusion and thus on transport properties.

\section{Quantum transport}

Within the Kubo-Greenwood formalism we compute the conductivity $\sigma(E_F)$ versus 
the Fermi energy $E_F$ using the real space method 
developped by Mayou, Khanna, Roche and Triozon,\cite{Mayou88,Mayou95,Roche97,Roche99,Triozon02} in the famework of the Relaxation Time Approximation (RTA) to account\cite{Trambly13} effects of inelastic scatterers due to electron-phonon interactions (see Appendix \ref{Sec_Kubo}).
Elastic scattering events due to local defects (vacant atoms) are included in the Hamiltonian itself in a large unit cell containing more than 10$^7$ atoms with boundary periodic conditions.

\subsection{High temperature conductivity}
\label{Sec_transpHT}

\begin{figure}

\includegraphics[width=6cm]{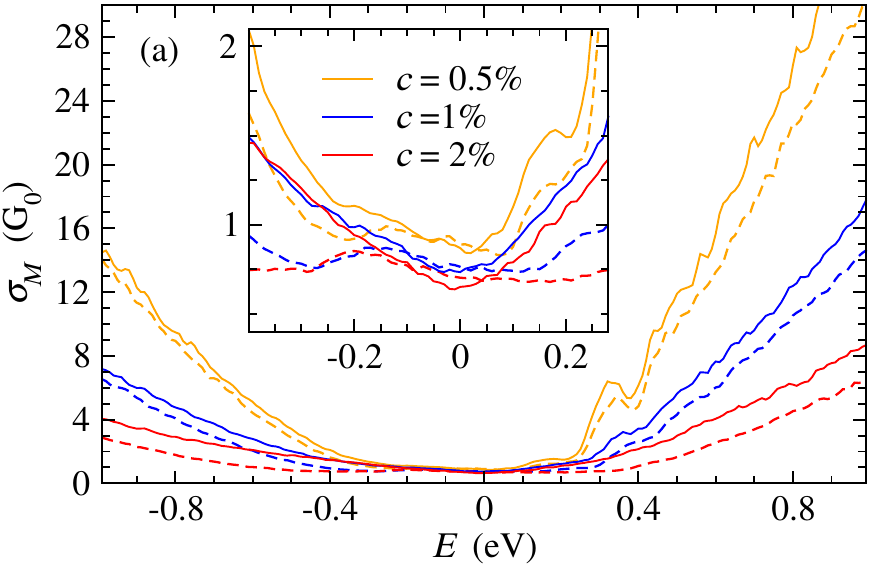} 

\includegraphics[width=6cm]{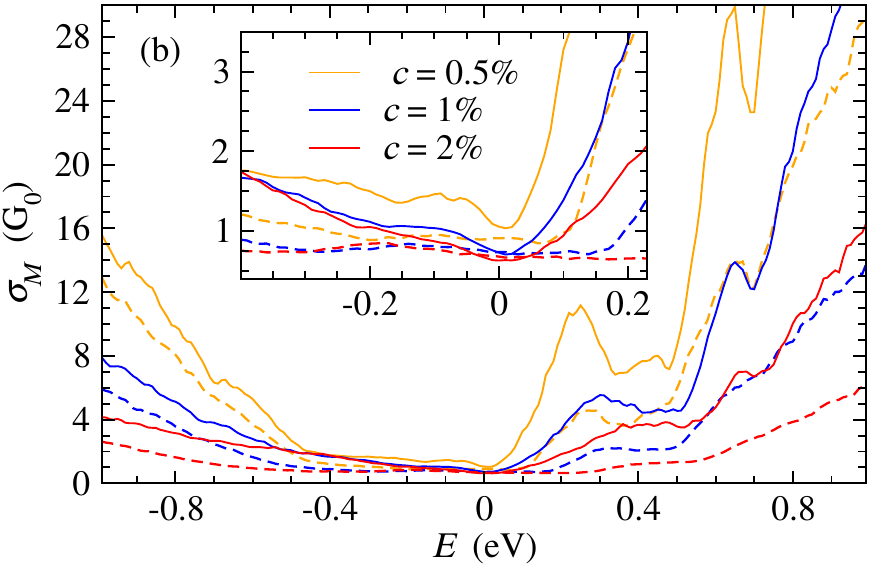}

\includegraphics[width=6cm]{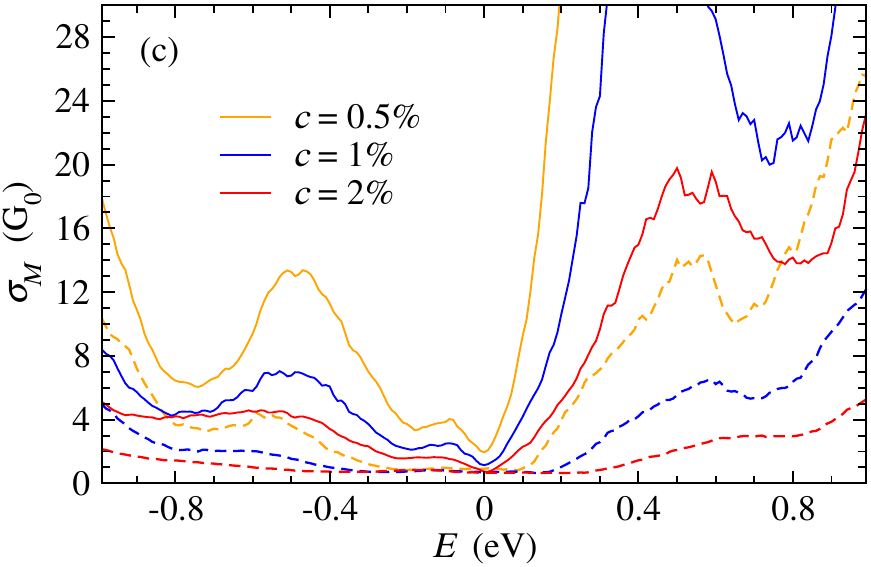}

\caption{ \label{Fig_SigmaM_fE}
(color online)
Microscopic conductivity $\sigma_M$ in (a) (12,13) tBLG, (b) (6,7) tBLG, (c) (3,4) tBLG, for the two cases:
(Full line) with vacancies in layer 2
and (dashed line) with vacancies in both layers.
$c$ is the concentration of vacancies with respect to the total number of atom in tBLG. Inserts: $\sigma_M$ around the Dirac energy $E_D=0$.
}
\end{figure}

\begin{figure}

\includegraphics[width=4.1cm]{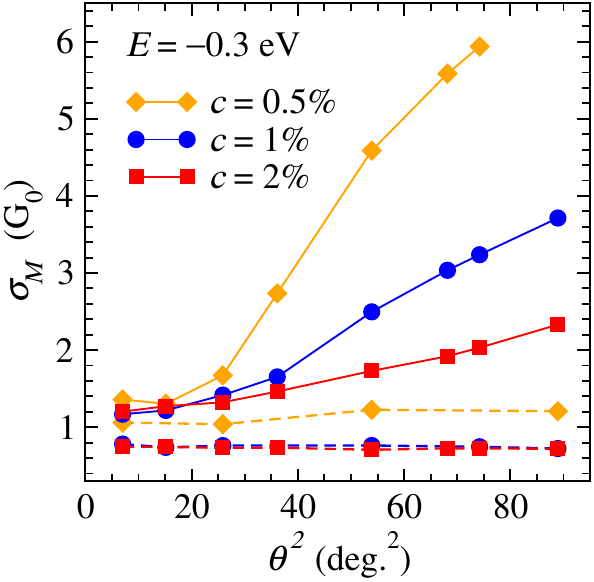} 
~ \includegraphics[width=4.1cm]{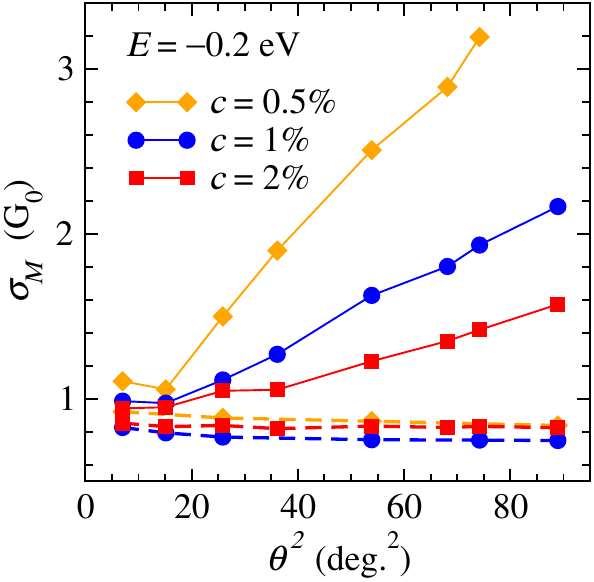} 

\vspace{.1cm}
\includegraphics[width=4.1cm]{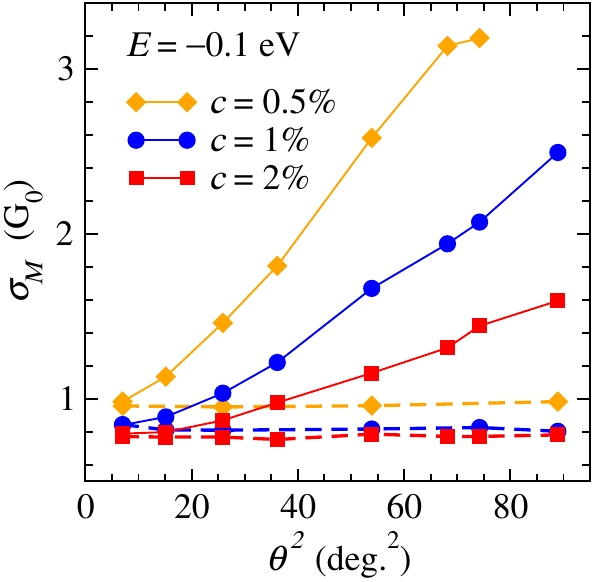} 
~ \includegraphics[width=4.1cm]{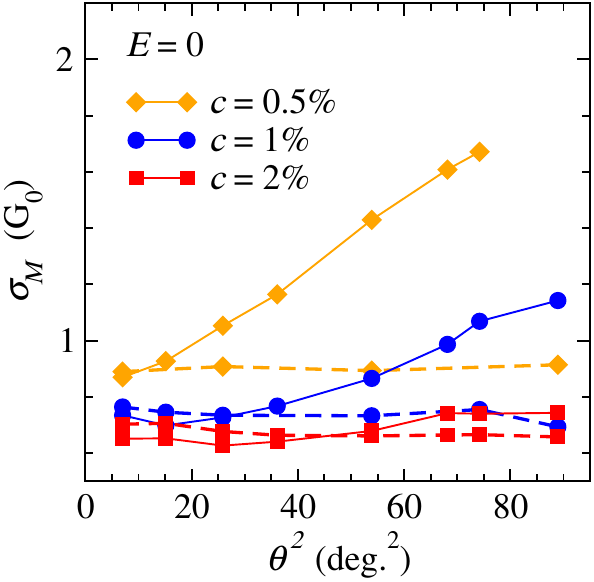}

\vspace{.1cm}
\includegraphics[width=4.1cm]{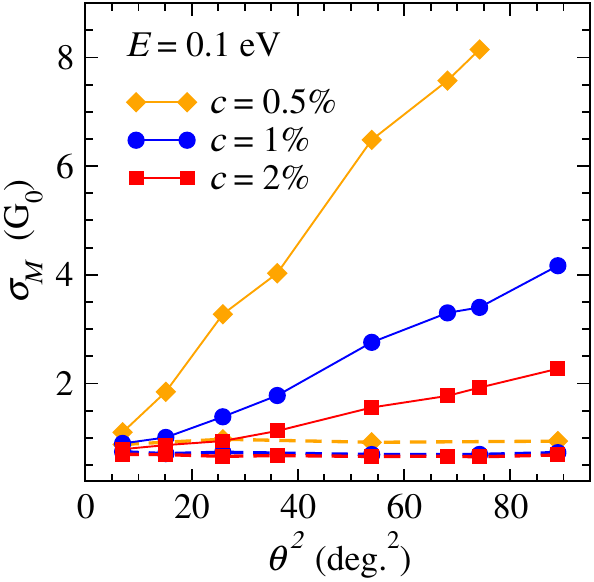} 
~ \includegraphics[width=4.1cm]{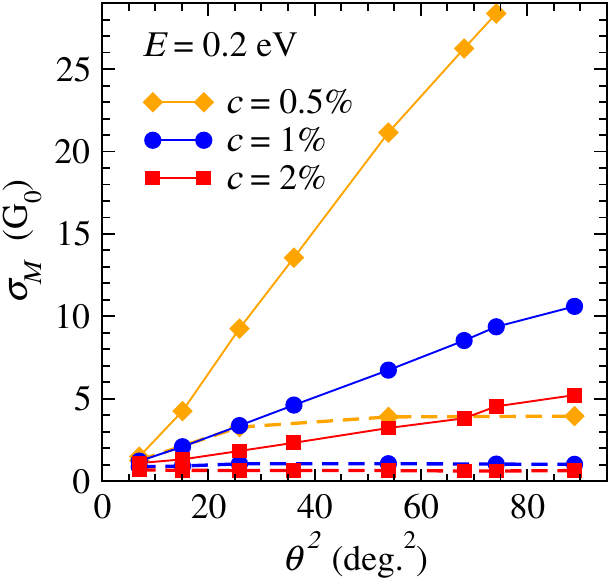}

\caption{ \label{Fig_SigmaM_fTheta}
(color online)
Microscopic conductivity $\sigma_M$ in tBLG versus rotated angle $\theta^2$ for energy values $E$:
(Full line) with vacancies in layer 2, (Dashed line) with vacancies in both layers. 
$c$ is the concentration of vacancies with respect to the total number of atom in tBLG. $E_D=0$.
}
\end{figure}

We first consider the high temperature case (or room temperature case) where the inelastic scattering time $\tau_i$ is close to the elastic scattering time $\tau_e$ due to static defects.
In that case, the dc-conductivity is called \emph{microscopic conductivity}, $\sigma_M$, because it takes into account quantum interference effects occurring during time less or equal to $\tau_e \simeq \tau_i$.  
$\sigma_M$ is close to semi-classical conductivity that does not take into account the quantum corrections due to multiple scattering effects. 
Typically, this quantity represents a room temperature conductivity when multiple scattering effects are destroyed by dephasing due to the electron-phonon interactions.
In Fig. \ref{Fig_SigmaM_fE},  $\sigma_M(E)$ is shown for three tBLG $(12,13)$, $(6,7)$ and $(3,4)$, with rotated angle $\theta$ equal to 2.656$^\circ$, 5.086$^\circ$ and 9.430$^\circ$, respectively, and
in Fig. \ref{Fig_SigmaM_fTheta}, $\sigma_M(\theta^2)$ is shown for different energy values close to the Dirac energy $E_{D}$.

For vacancies distribution $(i)$ --$ie$ vacancies randomly distruted in the two layers--, $\sigma_M(E)$ is almost independent of $\theta$ value. When vacancies concentration $c$ is large (Fig. \ref{Fig_SigmaM_fTheta}, $c = 1\%$ and $2\%$) behavior is similar to that of MLG and $\sigma_M \simeq 2 \sigma_{M,MLG}$, 
where $\sigma_{M,MLG}$ is the conductivity for MLG,\cite{Trambly13} 
$\sigma_{M,MLG} \simeq 0.6$\,G$_0$, with G$_0 = 2e^2/h$.
$\sigma_{M,MLG}$ reaches to the well known universal minimum of the conductivity so-called conductivity 
``plateau'' --independent of defect concentration-- at energies around $E_{\rm D}$.\cite{Castro09_RevModPhys} 
For smaller concentration (Fig. \ref{Fig_SigmaM_fTheta}, $c = 0.5\%$),  $\sigma_{M}$ increases when the concentration $c$ increases. 
These two regimes are similar to the one found in AB Bernal bilayer graphene.\cite{Missaoui17} 
Roughly speaking, 
for large $c$ values,  
the elastic mean free path $L_e$ in MLG (see Fig. 4(a) in Ref. \onlinecite{Missaoui17}) is smaller than the average traveling distance\cite{Missaoui17} $l_1$ in a layer between two interlayer hoppings of the charge carriers, and thus carriers behaves like in MLG. Whereas for small $c$ values, $L_e > l_1$ and thus interlayer hopping are involved in the diffusive regime and BLG conductivity properties are different that MLG ones.   

For vacancies distribution $(ii)$ --$ie$ vacancies randomly distributed in layer 2--,
and large rotated angle (Fig. \ref{Fig_SigmaM_fE}(c)), conductivity is larger than in the first case $(i)$. 
Indeed for large $\theta$, typically $\theta > 10^\circ$, eigenstates are located mainly in one layer (``\emph{decoupled}'' layers)\cite{Trambly10,Trambly16} 
and thus conductivity of the bilayer is the sum of the conductivity of two almost independent layers, 
\begin{equation}
\sigma_M \simeq \sigma_{M,1} + \sigma_{M,2},
\label{eq_sigma_M}
\end{equation}
corresponding to conductivity of layer 1 and 2, respectively.  
Conductivity of layer with defects is close to MLG conductivity $\sigma_{M,2} \simeq  \sigma_{M,MLG}$ and conductivity of layer without defects $\sigma_{M,1}$
is affected by the presence of defects in layer 2. 
With increasing $\theta$, the eigenstates are more and more located on one layer, thus layers are more and more decoupled, and the $\sigma_{M,1}$ increases as layer 1 becomes more and more like a pristine MLG. Consequently the conductivity of the tBLG increases when $\theta$ increases. 
In theses cases numerical results (Figs. \ref{Fig_SigmaM_fTheta}) show that $\sigma_{M}$ increases as $\theta^2$.  

For small angles (Fig. \ref{Fig_SigmaM_fE}(a) and Fig. \ref{Fig_SigmaM_fTheta}), eigenstates are located almost equally on both layer for all energies around Dirac energy;\cite{Trambly16} therefore they are affected in a similar way by the two kinds of vacancies distributions $(i)$ and $(ii)$. Conductivity is thus very similar in the two cases.

The analytical model presented in Appendix Sec. \ref{Sec_lifetime}, allows to understand why $\sigma_M$ increases as $\theta^2$  when defects are located only in layer 2 (cases $(ii)$). 
From Einstein conductivity formula, conductivity in layer $p$, $p=1$, $2$, is
\begin{equation}
\sigma_{M,p}(E) = e^2 \rho_p(E) v^2 \tau_p ,
\label{eq_sigma_M12} 
\end{equation}
where $\rho_p$ and $\tau_p$ are the average DOS in layer $p$ and the average elastic scattering time in layer $p$, respectively. 
For energy values in the plateau of conductivity arround $E_{\rm D}$, the layer 2 --with defects-- has a conductivity close to universal minimum of MLG,\cite{Trambly13}  
$\sigma_{M,2}(E) \simeq \sigma_{M,MLG}$, thus from equations (\ref{eq_sigma_M}) and (\ref{eq_sigma_M12}), the conductivity in the bilayer 
is
\begin{equation}
\sigma_{M}(E) \simeq \sigma_{M,MLG} \left( 1 + \frac{\rho_1(E)}{\rho_2(E)} \frac{ \tau_1}{\tau_2} \right),
\label{eq_sigma_bi_FTau}
\end{equation}
where the ratio between scattering times can be estimated from the formula (\ref{eq_rapport_tau}) obtained in the Appendix. Thus,
\begin{equation}
\sigma_{M}(E) \simeq \sigma_{M,MLG} \left( 1 + \frac{\rho_1(E)}{\rho_2(E)} \frac{\theta^2}{\theta_0^2}  \right),
\label{eq_sigma_bi_Ftheta}
\end{equation}
with $\theta_0 $ related to $\theta_1 $ (Appendix equation (\ref{theta1})),
\begin{equation}
\theta_0 = \sqrt{3} \theta_1 ,
\label{theta0}
\end{equation}
 i.e.  $\theta_0\simeq 2 ^\circ$ (Appendix Sec. \ref{Sec_velocity_renorm}).
Since $\rho_1(E)$ and $\rho_2(E)$ are different (Fig. \ref{Fig_tDOS_P_Va}) and depend on the energy values and the defect concentration $c$, the slope of $\sigma_M$ versus $\theta^2$ also depends  on $E$ and $c$ (Fig. \ref{Fig_SigmaM_fTheta}).

\subsection{Low temperature conductivity}
\label{Sec_transpLT}

\begin{figure}

\includegraphics[width=4.29cm]{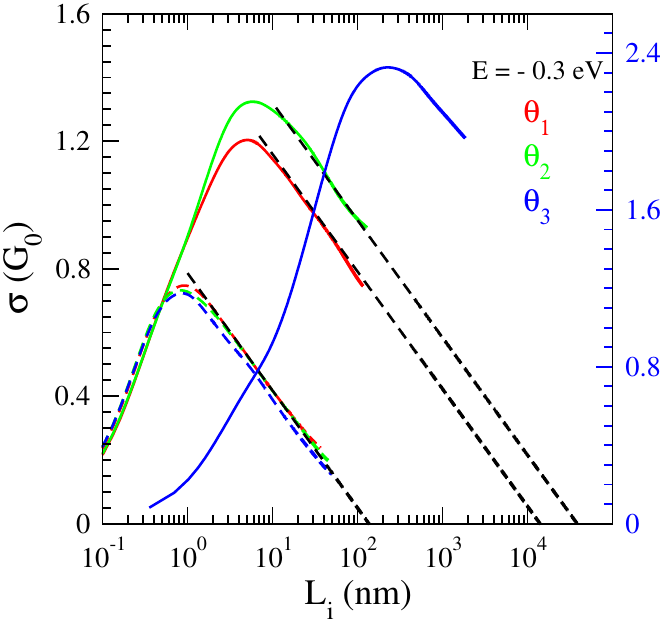} 
~ \includegraphics[width=4.1cm]{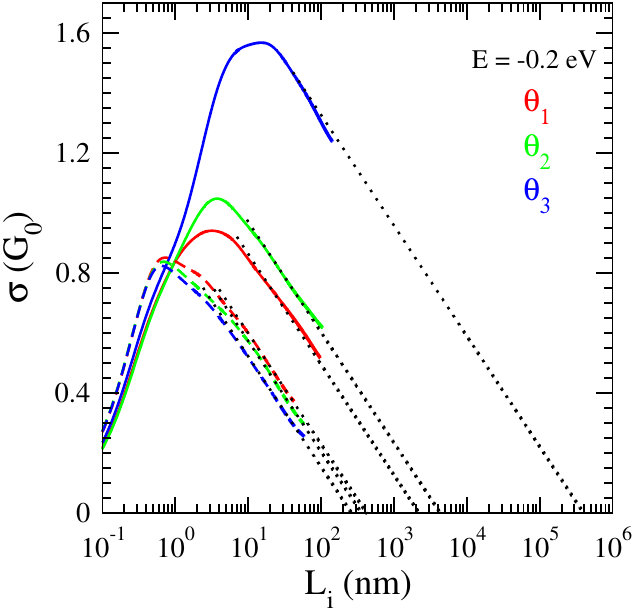} 

\vspace{.1cm}
\includegraphics[width=4.25cm]{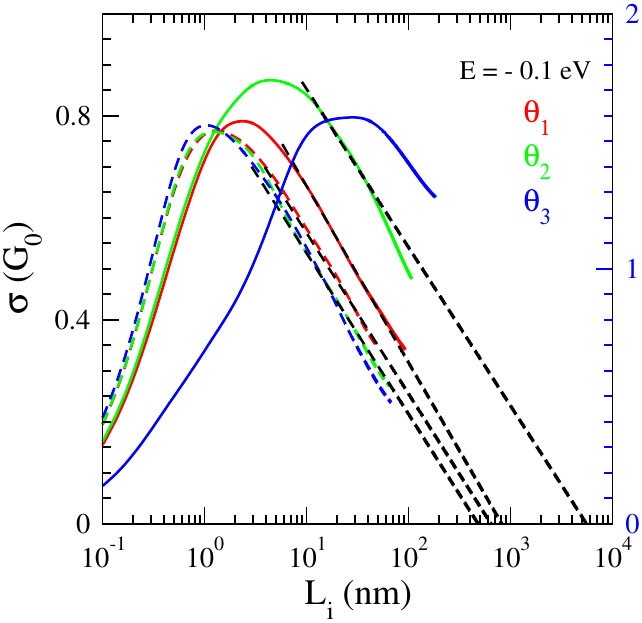} 
~ \includegraphics[width=4.1cm]{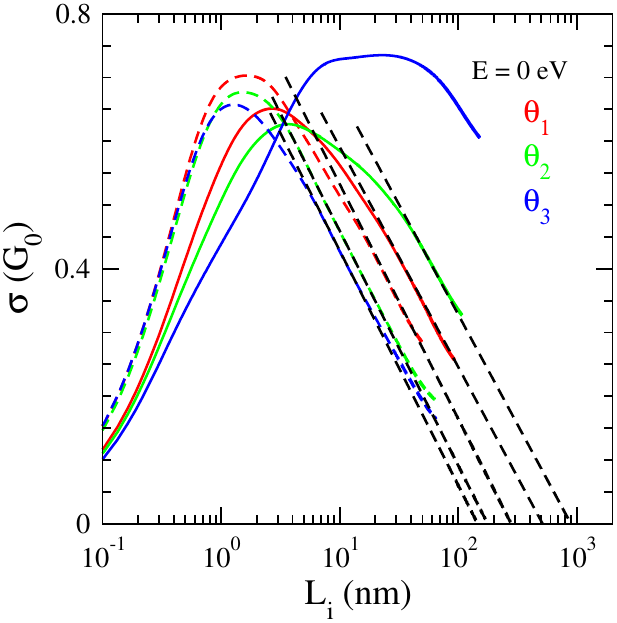}

\vspace{.1cm}
\includegraphics[width=4.1cm]{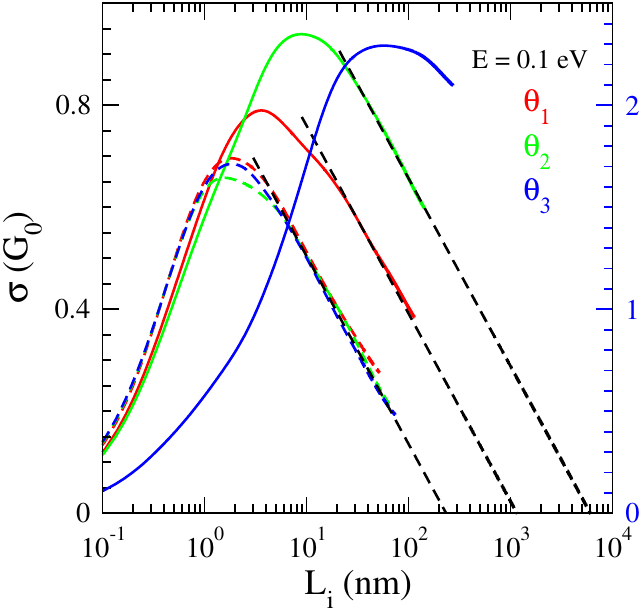} 
~ \includegraphics[width=4.1cm]{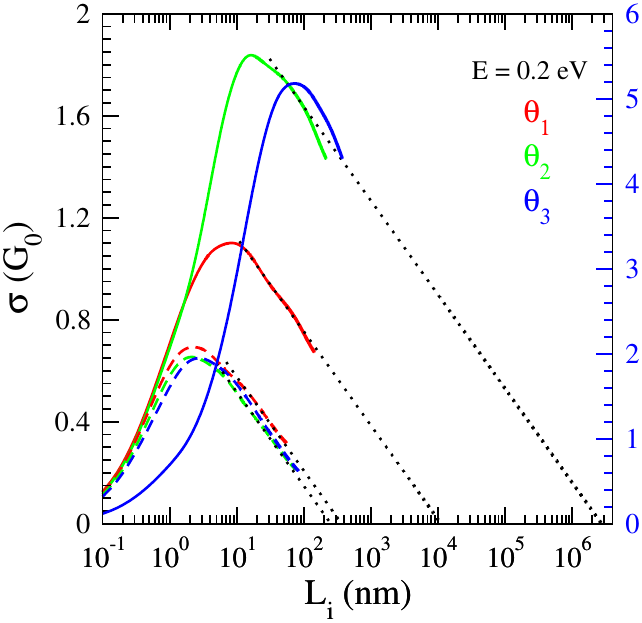}

\caption{ \label{Fig_Sigma_fLi}
(color online)
Conductivity in bilayer versus inelastic scattering $L_i$, at the energies $E$, for concentration $c=2$\% of vacancies with respect to the total number of atom in bilayer: 
($\theta_1 =  2.656{\rm ^o}$) (12,13) tBLG, 
($\theta_2 =  5.086{\rm ^o}$) (6,7) tBLG,
($\theta_3 =  9.430{\rm ^o}$) (3,4) tBLG.
(line) with vacancies in layer 2, (dashed line) with vacancies in both layers. 
%\textcolor{red}{Des calculs sont en cours pour avoir des valeurs de $\sigma$ pour $L_i$ plus grands.}
For (3,4) tBLG ($\theta_3 =  9.430{\rm ^o}$) the localization regim appears at very large times for which very time consuming calculations are necessary; 
that is why this regime is only roughly estimated by extrapolation. 
}
\end{figure}

\begin{figure}

\includegraphics[width=6cm]{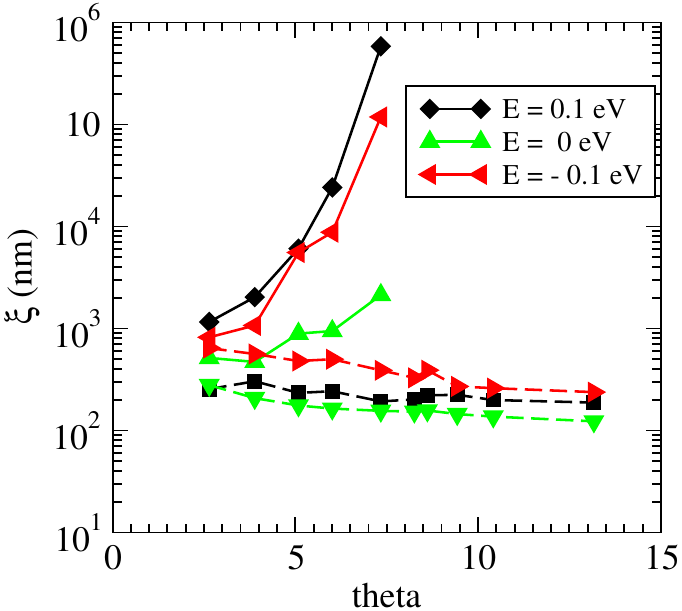} 

\caption{ \label{Fig_Loc_ftheta}
(color online)
Localization length versus angle  $\theta$, at the energies $E$, for concentration $c=2$\% of vacancies with respect to the total number of atom in bilayer:
(line) with vacancies in layer 2, (dashed line) with vacancies in both layers.
}
\end{figure}

In the low temperature limit, inelastic scattering time $\tau_i$ is larger than elastic scattering time $\tau_e$, and multiple scattering effects may reduced the conductivity with respect to microscopic conductivity $\sigma_m$. The average inelastic length $L_i$ thus satisfies  $L_i \gg L_e$ and $L_i \gg l_1$. $\tau_i$ and $L_i$ increase when temperature decreases. 
To evaluate this effect we compute\cite{Trambly13,Missaoui17} the conductivity $\sigma$ versus $L_i$ at every energy $E$ (Fig. \ref{Fig_Sigma_fLi}) for the two vacancy distribution cases: ($i$)  in the two layers and ($ii$)  in layer 2. 
As expected in disordered 2D systems,\cite{Lee85} for large $L_i$, $\sigma(L_i)$ follows a linear variation with the logarithm of $L_i$, like in the case of monolayer graphene \cite{Trambly13,Trambly11} and Bernal bilayer graphene,\cite{Missaoui17}
\begin{equation}
\sigma(E,L_i) = \sigma_0 - \alpha {\rm G}_0 \ln (L_i) ,
\end{equation}
where $\sigma_0$ is a constant depending on $\sigma_M$ and $L_e$, and the slope $\alpha$ is almost independent on energy $E$, the defect concentration and the repartition of the defects (in one layer or in both layers). 
From numerical results one obtains $\alpha \simeq 0.32$ 
which is close to monolayer value\cite{Trambly13} and Bernal bilayer value.\cite{Missaoui17}

Localization length $\xi$ can be estimated from the equation $\sigma(L_i = \xi) = 0$ and the linear extrapolation of $\sigma$ versus $\log L_i$ at large $L_i$ (see dashed lines Fig. \ref{Fig_Sigma_fLi}). $\xi$ versus $\theta$ for various energies in the plateau of conductivity are shown in Fig. \ref{Fig_Loc_ftheta}. As $\sigma_M$, $\xi$ is almost independent of $\theta$ when defects are located in both layers, but $\xi$ increases strongly when defects are located in one layer only.

\section*{Conclusion}

We have presented a numerical study of the local electronic density of states (LDOS) and the conductivity in pristine and covalently functionalized twisted graphene bilayers (tBLG), with an angle of rotation $\theta > 2 ^\circ$. 
Those results are understood using a perturbative analytical model described in the Appendixes. 
The atomic structure in Moir\'e induces  a strong modulation in the LDOS between AA stacking areas and AB stacking areas, which varies as $\theta^{-2}$ following a simple analytic expression. 
We show that disorder breaks the interlayer effective coupling due to Moir\'e pattern. 
Therefore when defects are randomly distributed in both layer, the conductivity $\sigma_M$ is almost independent of $\theta$, whereas $\sigma_M \sim \theta^2$ when defects are randomly distributed in one layer only. Such a non-symmetric distribution of defects may often occur in experimental situation because of the effect of substrate, adatoms or admolecules. Finally the quantum correction to the conductivity are computed and localization length is calculated versus $\theta$.

\section{Acknowledgment}
The authors wish to thank C. Berger, W. A. de Heer, P. Mallet and J.-Y. Veuillen, T. Le Quang, V. Renard, C. Chapelier
for fruitful discussions.
TB calculations have been performed 
at the Centre de Calculs (CDC),
Universit\'e de Cergy-Pontoise.
We thank Y. Costes and Baptiste Mary, CDC, for computing assistance. 
DFT calculations was performed using HPC resources from GENCI-IDRIS (Grants A0060910784 and 
A0060907655).
This work was supported by the ANR project J2D (ANR-15-CE24-0017) and the Paris//Seine excellence initiative (grant 2019-055-C01-A0).

\appendix

\section{Kubo-Greenwood conductivity}
\label{Sec_Kubo}

In Kubo-Greenwood approach for transport properties, 
the quantum diffusion $D$, is computed by using the polynomial expansion of the average square spreading, $\Delta X^{2}$, for charge carriers.
This method, developed by Mayou, Khanna, Roche and Triozon,\cite{Mayou88,Mayou95,Roche97,Roche99,Triozon02}  
allows very efficient numerical calculations by recursion in real-space that take into account all quantum effects.
%It has been used to studies quantum transport in disordered graphene \cite{Lherbier08,Lherbier11,Leconte11,Leconte11b,Roche12,Roche13,Trambly13}  and chemically doped graphene \cite{Lherbier08b,Leconte10}. 
Static defects are included directly in the structural modelisation of the system and they are randomly distributed on a supercell containing up to $10^7$ Carbon atoms. 
%This corresponds to typical sizes of about one micrometer square which allows to study systems with inelastic mean-free length of the order of few hundreds nanometers.
Inelastic scattering is computed \cite{Trambly13} within the
Relaxation Time Approximation (RTA) including an inelastic scattering time $\tau_i$ beyond which the propagation becomes diffusive due to the destruction of coherence by inelastic processes. 
One finally get the Einstein conductivity  formula, \cite{Trambly13}
\begin{equation}
\sigma(E_F,\tau_i) = e^2 \rho(E_F) D(E_F,\tau_i),
\label{eq_einstein}
\end{equation}
where $E_F$ is the Fermi level,
$D(E,\tau_i)$ is the diffusivity (diffusion coefficient at energy $E$ and inelastic scattering time $\tau_i$),
\begin{equation}
D(E,\tau_i) = \frac{L_i^2(E,\tau_i)}{2 \tau_i},
\end{equation}
$\rho(E)$ is the density of states (DOS) and $L_i(E,\tau_i)$ is the inelastic mean-free path. $L_i(E,\tau_i)$ is the typical distance of propagation during the time interval $\tau_i$ for electrons at energy $E$, 
\begin{equation}
L_{i}^{2}(E,\tau_{i})=\frac{1}{\tau_{i}} \int_0^\infty \! \Delta X^{2}(E,t)\,{\rm e}^{-t/\tau_{i}} \, dt .
\end{equation}
%The average square spreading $\Delta X (E,t)$ are energy $E$ and time $t$, is computed by a real-space method to determine the evolution of wave packets \cite{Mayou88,Mayou95,Roche97,Triozon02}. 
%Our calculations are performed on sample containing up to $10^7$ atoms, containing static defects (vacancies). 
%That corresponds to typical size of about one micrometer square which allows to study systems with inelastic mean-free length of the order of few hundreds nanometers.
%This method has been also used to study other many of defects in graphene based nanomaterials \cite{....}.
Without static defects (static disorder) the $L_i$ and $D$ goes to infinity when $\tau_i$ diverges. 
With statics defects, at every energy $E$, $\sigma(\tau_i)$ reaches a maximum value,  
\begin{equation}
\sigma_M(E_F,\tau_i) = e^2 n(E_F) \, {\rm Max}_{\tau_i} \left \{ D(E_F,\tau_i) \right \},
\label{eq_einstein_SigM}
\end{equation}
called \emph{microscopic conductivity}. 
$\sigma_M$ corresponds to the usual semi-classical approximation (semi-classical conductivity). 
This conductivity is typically the conductivity at room temperature, when inelastic scattering time $\tau_i$ (inelastic mean free path $L_i$) is close to elastic scattering time $\tau_e$ (elastic mean free path $L_e$), $\tau_e(E) = L_e(E) / v(E)$ and $L_e(E) = D_M(E) / 2 v(E)$, where $D_M(E)$ is the maximum value of $D(\tau_i)$ at energy $E$ and $v(E)$ the velocity at very small times (slope of $\Delta X(t)$).

For larger $\tau_i$ and $L_i$, $\tau_e \ll \tau_i$ and $L_e \ll L_i$, quantum interferences may result in a diffusive state, 
$D(\tau_i) \simeq D_M$, or a sub-diffusive state where $D$ decreases when $\tau_i$ and $L_i$ increase. 
For very large $L_i$, $L_i$ close to localization length $\xi$, the conductivity goes to zero. 
%These two last regimes ($L_e \ll L_i$, and  $L_i \simeq \xi$), which correspond to the low temperature regime,  are not discussed in the main paper.

\section{Tight-Binding Model}
\label{Sec_TB}

\subsection{Real space couplings}
\label{Sec_Hamiltonian_R}
\begin{figure}
\includegraphics[width=6.5cm]{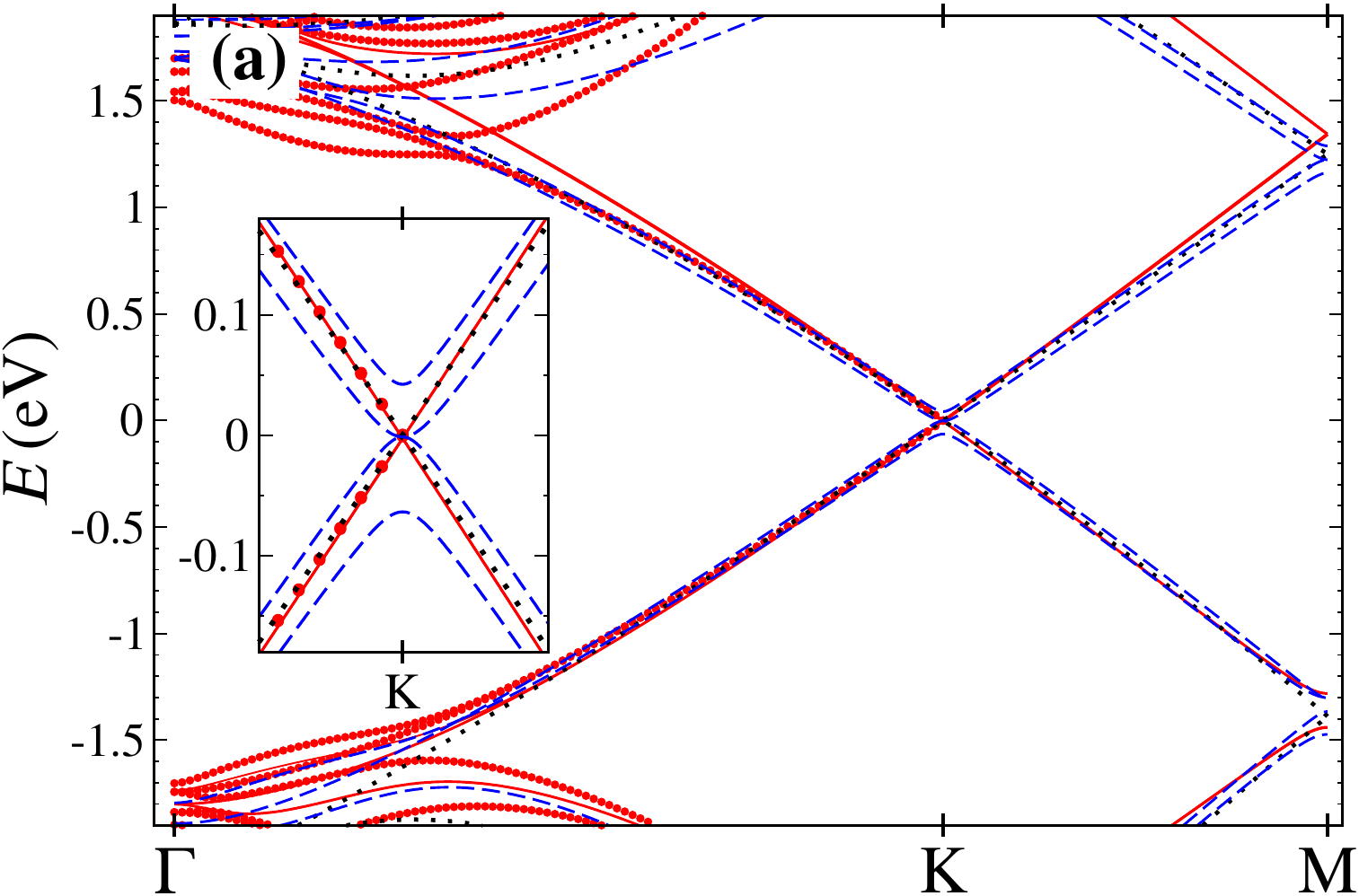}

\vskip .1cm
\includegraphics[width=6.5cm]{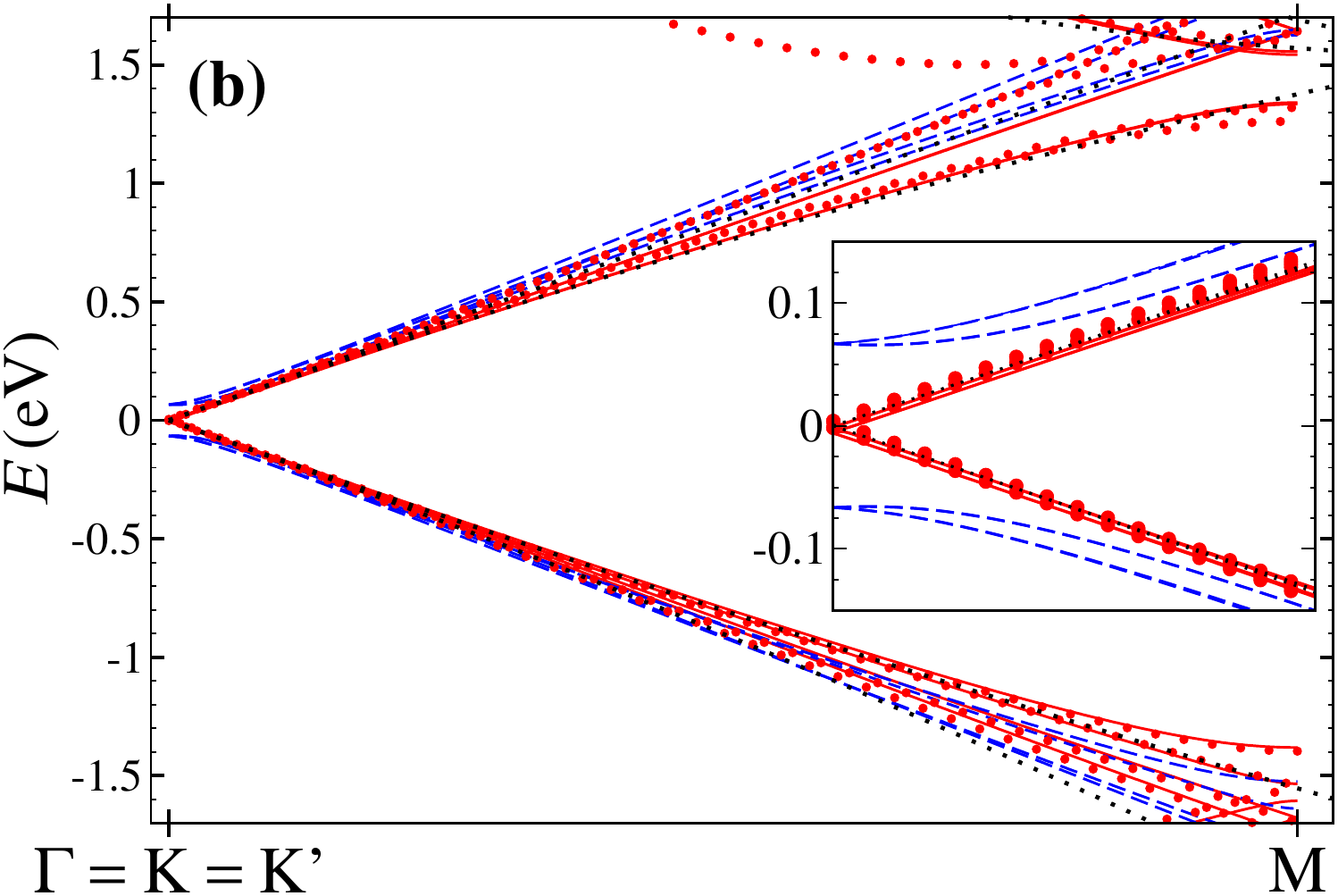}

\caption{(color on line) Band dispersion $E(\vec k)$: (red points) DFT calculation using VASP code (for details on the calculation see Ref. \onlinecite{Trambly12}), and 
(red lines) TB calculation,  for
(a)  (1,3) tBLG ($\theta = 32.20^o$), (b) (1,4) tBLG ($\theta = 38.21^o$), calculated with a large interlayer cutoff distance $r_c = 0.60$\,nm, whereas 
Blue dashed line shows TB bands with a too small $r_c$, $r_c = 0.34$\,nm. In the latter case a non-physical gap appears at energy $E_D = 0$.
Dot black line is for MLG.
Insert: zoom of the bands around the K point. $E_D=0$. 
}
\label{Fig_graph_1-3_1-4}
\end{figure}

In the tight-binding (TB) scheme only $\rm p_z$ orbitals are taken into
account since we are interested in electronic states close to
the Fermi level. The TB model used in this paper is the same as in our previous work on twisted bilayers graphene \cite{Trambly10,Trambly12,Trambly16} and in AB Bernal bilayer 
graphene.\cite{Missaoui17,Missaoui18}
The Hamiltonian has the form,
\begin{equation}
{H} =\sum_{i} \epsilon_{i}|i\rangle\langle i| + \sum_{(i,j)}t_{ij}|i\rangle\langle j| ,
\end{equation}
where $i$ is the p$_z$ orbital located at $\vec{r}_{i}$ with an on-site energy $\epsilon_{i}$, and  
the sum runs over all neighboring $i$, $j$ sites. $t_{ij}$ is the hopping element matrix between site $i$ and site $j$, computed from the usual Slater-Koster parameters as given in Ref \onlinecite{Trambly12}. 
Since the layers are rotated, interlayer neighbors are
not on top of each other (as it is the case for the Bernal AB
stacking). Therefore, the interlayer hopping terms are then not restricted to
$\rm pp\sigma$ terms but $\rm pp\pi $ terms have also to be introduced.\cite{Trambly10,Trambly12} 
Moreover hopping terms are not restricted to first neighbouring orbitals and they decrease exponentially with the interatomic distance. 
A cutoff distance $r_c$ is introduced which must be large enough so that the results do not depend on it. 
We have checked that $r_c  = 0.6$\,nm is enough. 
%Note that if $r_c$ is too small, a non-physical small gap may appeared at the Dirac energy as shown in Fig. \ref{Fig_graph_1-3_1-4}.  
For small $r_c$ values, small gap may appeared at the Dirac energy as shown in Fig. \ref{Fig_graph_1-3_1-4}. 
Several studies \cite{Shallcross08,Mele10,Sboychakov15,Rozhkov17} has shown that this small gap comes from
non-zero matrix element coupling electron 
states in equivalent Dirac cones for some superstructures with small number of atoms in the cell of tBLG. 

\begin{figure}[!h]
\centering
%\vspace{2mm}
%\scalebox{0.5}{\includegraphics{coupling.png}}
\scalebox{0.45}{\includegraphics{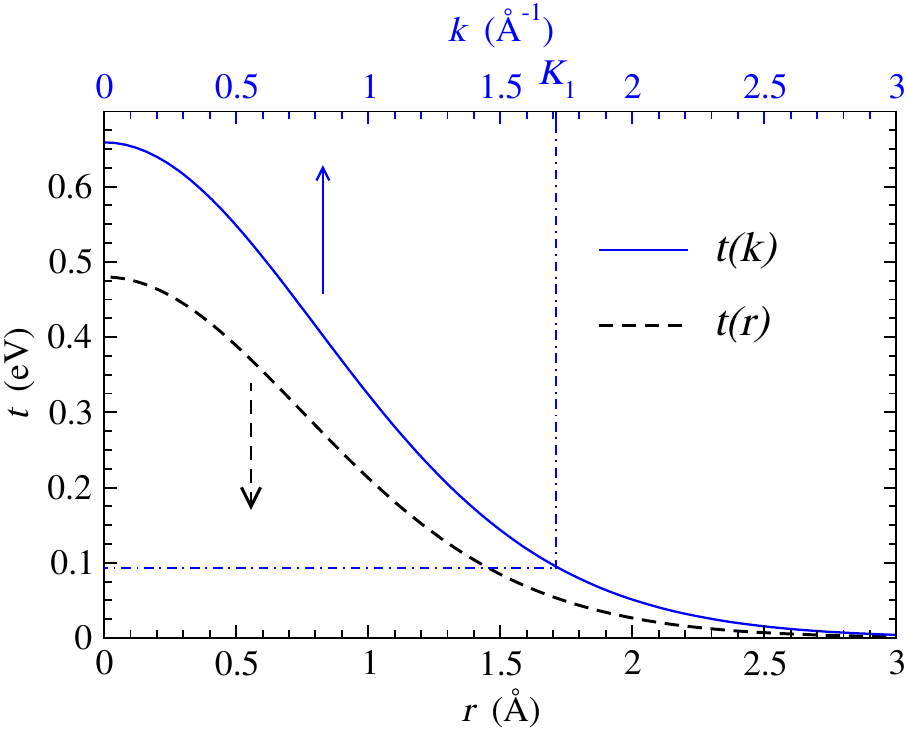}}       
\caption{Modulus of the interlayer coupling $t$ versus in-layer distance $r$ and modulus $k$ of the wave-vector, calculated from the Tight-binding model described in Ref. \onlinecite{Trambly12}.
      }
\label{fig:coupling}
\end{figure}

The matrix element of the interlayer Hamiltonian $H_c $ between one orbital at $\vec{r}$ in layer 1 and one orbital at $\vec{r}\,'$ in layer 2 is given by 
\begin{equation}   
{\langle} \vec{r'} | {H_c} |  \vec{r}{ \rangle}=  H_{c}( | \vec{r}- \vec{r}\,'|) .
\end{equation} 
Note that $H_{c}( \vec{r}- \vec{r}\,')$ is real and depends only on the modulus  $| \vec{r}- \vec{r}\,' |$. $H_{c}(| \vec{r}|)$ is maximum at zero distance i.e. when the two orbitals are aligned perpendicularly to the two layers. 
The hopping integral between the two orbitals decreases when their distance increases. The Fourier transform which will be essential in the following is also real and depends only on the modulus of the wave vector. From Fourier transformation we write
\begin{equation} 
 H_{c}(\vec{r})=\displaystyle  \int{{\widetilde{H}}_{c}(\vec{k}) \, \,  {\rm e}^{i\vec{k} \cdot \vec{r}} \,    {d^2}\vec{k} } \, ,
\end{equation} 
and
\begin{equation} 
  \widetilde{H}_{c}(\vec{k})= \frac{1}{(2\pi)^2} \int H_{c}(\vec{r}) \,   {\rm e}^{-i\vec{k} \cdot \vec{r}} \, {d^2}\vec{r} .
\end{equation} 
Here also the coupling $ \widetilde{H}_{c}(\vec{k})$ decreases when $|\vec{k}|$ increases. We shall see below that the largest value of $ \widetilde{H}_{c}(\vec{k})$ is for $|\vec{k}|$
close to the modulus of a Dirac point which is represented by $K_D$ in Fig. \ref{fig:coupling}. 

\subsection{Interlayer Coupling  between Bloch states}
\label{Sec_InterlayerCoupling}

We want to compute the coupling between two Bloch states of layer 1 and layer 2. 
Each graphene layer is a honeycomb lattice with two atoms, atoms A and atoms B, in its unit cell.
Let us consider normalized Bloch states made of atomic $\rm p_z$ orbitals A or B in layer $\alpha$,
$\alpha = 1 $ or $2$,   
\begin{equation}
 \label{bloch1}
 {\vert A \vec{k} \rangle}_\alpha =    \frac{1}{\sqrt{N}}      \sum\limits_{\vec{R}} e^{ i \vec{k} \cdot \vec{R_A}} {\vert {A  \vec R} \rangle}_\alpha  ,
\end{equation}
\begin{equation}
 \label{bloch2}
 {\vert B \vec{k} \rangle}_\alpha =      \frac{1}{\sqrt{N}}   \sum\limits_{\vec{R}} e^{ i \vec{k} \cdot \vec{R_B}} {\vert {B  \vec R} \rangle}_\alpha  ,
\end{equation}
where $N$ is the number of unit cells of the crystal and the summation is performed on all cells of crystal ($\vec{R}$).
In the following A  or  B are indicated by $ \varepsilon $   according to the following convention,   
\begin{equation} \varepsilon= \begin{cases}
 {\rm A} & {\rm for~A~atom}  \\
 {\rm B} & {\rm for~B~atom}   \end{cases} 
 \end{equation}
\begin{equation} \alpha= \begin{cases}
 1  & \mbox{lower layer}  \\
 2  & \mbox{upper layer}   \end{cases} 
 \end{equation}        
%We want to compute the matrix element of the Hamiltonian between  two normalized Bloch states $  {| {\varepsilon}  \vec{k} \rangle}_1  $  and $  {| {\varepsilon}'  \vec{{k'}} \rangle}_2  $ :
%\begin{equation}  
%\ {H_c}(\varepsilon,{\varepsilon}' ; \vec{k} , \vec{k'}) = \langle \vec{k'} {\varepsilon}' | {H_c} | {\varepsilon}  \vec{k} \rangle = \ {H_c}( \vec{k'} {\varepsilon}' ; \vec{k}  \varepsilon) \, .
%\end{equation}
%As $ {H_c} $ is Hermitian, 
%\begin{equation}  {_2}\langle \vec{k'} {\varepsilon}' | {H_c} | {\varepsilon}  \vec{k}{{ \rangle}}_1 =  {\left( {_1}\langle {\varepsilon}  \vec{k} | {H_c} | \langle  {\varepsilon}' \vec{k'} {{ \rangle}_2} \right)}^{\ast} \, ,
%\end{equation}
%one obtains
%\begin{equation}
%  \ {H_c}(\varepsilon,{\varepsilon}' ; \vec{k} , \vec{k'}) =  \sum\limits_{\vec{R} , \vec{R'}} e^{ i (\vec{k} \cdot \vec{R} - \vec{k'} \cdot \vec{R'})}        {_1}\langle \vec{R'} {\varepsilon}' | {H_c} | {\varepsilon}  \vec{R}{{ \rangle}}_2   .
% \end{equation}
The positions of the atoms in layer 1 are,
\begin{equation} \begin{cases}
  \vec{r}_{\varepsilon\vec{R}} = \vec{R}  & \mbox{if $ \varepsilon = A $}  \\
 \vec{r}_{\varepsilon\vec{R}} = \vec{R}+ \vec{u} & \mbox{if $ \varepsilon = B $}  
\end{cases} \end{equation}
and in layer 2,
\begin{equation} \begin{cases}
 \vec{r}\,'_{{\varepsilon}'\vec{R}\,'} = \vec{R}\,'  & \mbox{if $ {\varepsilon}' = A $}  \\ 
 \vec{r'}_{{\varepsilon}'\vec{R}\,'} = \vec{R'}+ \vec{{u}}\,'  & \mbox{if $ {\varepsilon}' = B $}  
 \end{cases} \end{equation}
where $\vec{u}$ and $\vec{u}\,'$ are  vectors connecting the two atoms in the unit cells, i.e. $A$ and $B$ atoms in layer 1 and $A'$ and $B'$ atoms in layer 2, respectively.
Writing
\begin{equation}
H_{c} |\varepsilon \, \vec{k} {\rangle}_2 =\sum\limits_{i} t({\varepsilon}_i \, {\vec{k}}_i \,, \varepsilon\,\vec{k})\,\,| {\varepsilon}_i \, {\vec{k}}_i {\rangle}_1 ,
\end{equation}
where $ t({\varepsilon}_i \, {\vec{k}}_i \,, \varepsilon\,\vec{k}) \,  \equiv t_i   $ is the transfer matrix element, we find a selection rule such that 
\begin{equation}
\displaystyle {  \vec{k} + \vec{K}_r = \vec{k}\,'+\vec{K}\,'_r   },
\label{Selection}
\end{equation}
where $\vec{K}_r    $ and $\vec{K}_r' $ are vectors of reciprocal lattices.
This means that interlayer coupling Hamiltonian $H_c$  couples the upper state $ |\varepsilon \, \vec{k} {\rangle}_2 $  to lower state $ |\varepsilon \, \vec{k} {\rangle}_1 $ only if the selection rule equation (\ref{Selection}) is obeyed.

Finally for $\vec{k}_i =\vec{k} + \vec{K}_r =\vec{k}\,' \,({\rm mod}\,\, \vec{K}_r') $, 
we derive a formula for coupling matrix, after some calculations,\cite{Omid_these} we switch to the following expression of the Hamiltonian,
\begin{equation}
\displaystyle  t_i(\vec{k} +\vec{K_r})  =\displaystyle  \frac{4\pi^2}{S}\,{\widetilde{H}_c}( \vec{k} + \vec{K_r} ) \, {\rm e}^{i\, ( \vec{k} +\vec{K_r} ) \cdot({\varepsilon}' \vec{u'} - \varepsilon\vec{u}+\vec{\Delta}) }.
\end{equation}
$S$ is the area of the unit cell, $\vec{\Delta}$ is the translation between the two layers. However this translation of the two layers just translates the overall Moir\'{e}  pattern and can be set to zero without loss of generality.

By symmetry of hopping term between two orbitals, coupling depends only on the modulus of $\vec{k} + \vec{K_r}$ i.e $  \widetilde{H}_c( \vec{k} + \vec{K_r} )  \simeq \widetilde{H}_c(| \vec{K}_D + \vec{K_r}| )   $, 
in the vicinity of the Dirac point. 
The modulus of $t_i$ is represented in Fig. \ref{fig:coupling}.
One sees that the largest value of $ |t_i|$ is one that corresponds to the smallest possible value of $  \vec{k} + \vec{K}_r$. By careful examination it can be shown that for electronic states close to the Dirac point this minimum corresponds to the modulus of wave-vector in Dirac point ($K_D = ||\vec{K}_D|| \simeq 17.2$\,nm$^{-1}$). From Fig. \ref{fig:coupling}, it  is easy to deduce numerically the interlayer hopping term close Dirac is around $t_1 \simeq 0.12$\,eV.
All the other contributions are much smaller and will be neglected here.

Selecting only this contribution means that $\vec{K_r}$ is such that $\vec{k} + \vec{K_r}$ belongs to one of three equivalent valleys. Therefore a set of two Bloch states with a given wave vector  (equations (\ref{bloch1}) and (\ref{bloch2})) in one layer will be coupled to three sets of two Bloch states in other layer corresponding to three different wave vectors. This strongly simplifies the structure of Hamiltonian and the analytical calculations presented here. 

In the following we shall count the vectors $\vec{k}$ and $\vec{k}'$ from their respective Dirac point $\vec{K}_{D1}$ and $\vec{K}'_{D1}$. $\vec{K}'_{D1}$ is obtained from $\vec{K}_{D1}$ by a rotation of an angle $\theta$ around the vector $\vec{\zeta}$ which is perpendicular to the layers 1 and 2. Therefore one has
\begin{equation}
\displaystyle {\vec{k} = \delta \vec{ k} +\vec{K}_{D1}} ,
\label{defdK}
\end{equation}
\begin{equation}
\displaystyle {\vec{k}\,' = \delta \vec{k}\,' +\vec{K}'_{D1}} .
\label{defdK'}
\end{equation}
Finally one get for the selection rule
\begin{equation}
%\displaystyle 
\delta {\vec{k}\,' \simeq \delta \vec{k} - \theta \vec{\zeta} \times \vec{K}_{Di} },
\label{SRdK}
\end{equation}
where the indice $i$ takes the values $i=1,2,3$. $\vec{K}_{Di}$ and $\vec{K}'_{Di}$ are the three equivalent Dirac point in layer 1 and 2. $\vec{K}'_{Di}$ is obtained from $\vec{K}_{Di}$ by a rotation of an angle $\theta$ around the vector $\vec{\zeta}$.

\section{Effect of interlayer coupling}
\label{Sec_IC}

We consider a Layer 1 coupled to layer 2 which is rotated by an angle $\theta$ with respect to layer 1. If one considers the time evolution within layer 1 or more generally the restriction of the total Green's function to layer 1, the coupling to layer 2 amounts to the addition of an effective Hamiltonian or self-energy. From this effective Hamiltonian we shall get the velocity renormalization, the electron lifetime in layer 1 due to disorder in layer 2 and the modulation of the DOS close the charge neutrality point. The theory which is developed here is perturbative and assumes that the rotation angle $\theta$ is not too small. In particular we emphasize that the perturbation theory is valid for
\begin{equation}
{ \displaystyle  z,\, t,\, \Delta \ll \hbar v K_D \theta   },
\end{equation}
where $v$ is the monolayer velocity and $K_D = ||\vec{K_D}||$, $z$ is the energy of calculation, $t$ is the interlayer coupling ($t\simeq t_1 \simeq 0.12$\,eV, Sec. \ref{Sec_Hamiltonian_R}) and $\Delta$ is a possible difference in on-site energy between the two layers. The condition on $t$ implies that $\theta > \theta_1$ where
\begin{equation}
\theta_{1}=\frac{\sqrt{2}t}{\hbar v K_D }.
\label{theta1}
\end{equation}
The value of $\theta_{1}$ is close to $\theta_{1} \simeq 1^\circ$. The condition on $z$ implies that the current energy at which the quantities are calculated  is smaller than the typical energy of the Van Hove Singularities (VHS) which depends linearly on $\theta$. The difference in energy $\Delta$ of the two layers must also be smaller than the energy of the VHS.Note that the VHS have been clearly observed with STM experiments on twisted graphene bilayer.

\subsection{Effective one-plan Hamiltonian}
\label{Sec_effective H one plane}
 
We consider first a Bloch state in layer 1 with wave vector $\delta\vec{k_0}$. It can be coupled to a Bloch state $\delta \vec{k'}$ in layer 2 then propagates freely in layer 2 and is scattered again to a Bloch state in layer 1 with a wave vector $\delta \vec{k_f}$. Applying the selection rule (\ref{SRdK}) to each interlayer hopping term we find that $\vec{\delta k_f}$ and $\vec{\delta k_0}$ are related by 
 \begin{equation}
%\displaystyle 
{\delta \vec{k_f} \simeq \delta\vec{ k_0} - \theta \vec{\zeta} \times (\vec{K}_{Di} -\vec{K}_{Dj}) }.
\label{SRdK11}
\end{equation}
Therefore the coupling between layers 1 and 2 induces an effective coupling between Bloch states of layer 1 with the  selection rule (\ref{SRdK11}). Note that the indices $i$ and $j$ take the values $1$, $2$, $3$. 

When $i=j$, a Bloch state with $\delta \vec{k_0}$ is coupled only to the Bloch states with the same wave-vector $\delta \vec{k_f} = \delta \vec{k_0}$. This process gives a self-energy which renormalizes the energy of the  state of the single layer 1 (see below). 

When $i$ and $j$ are different then $\delta\vec{ k_f} $ and $\delta \vec{k_0}$ are different, 
\begin{equation}
%\displaystyle 
{\delta \vec{k_f} \simeq \delta \vec{k_0} +\vec{G_k}  }.
\label{SRdKG11}
\end{equation}
$\vec{G_k}=\theta \vec{\zeta} \times (\vec{K}_{Di} -\vec{K}_{Dj})$ is a reciprocal lattice vector of the Moir\'{e} lattice, 
where
$\vec{K}_{Di} -\vec{K}_{Dj}$ is a reciprocal lattice vector of graphene.
These  vectors takes six possible values, named $\vec{G_k}$ in the main text, that are vectors of the reciprocal lattice of the Moir\'{e} pattern. As we show below this coupling between Bloch states of different wave vector will create eigenstates with mixing of different oscillating components which leads to oscillations in the DOS with wave-vectors components $\vec{G_k}$ (see below). We note also that the coupling introduces only small spatial frequencies and in particular it does not connect states of the two non equivalent Dirac cones.

\subsection{Self-energy}
\label{Sec_SelfEnergy}

We are interested in the self-energy of coupling of states in layer 1 due to the coupling with states of layer 2. Indeed the real-part of self-energy $ \Re \sigma(z) $ is associated to modification of dispersion relation and will allow us to discuss velocity renormalization. The imaginary part of self-energy is associated to the electron lifetime. It will allow us to discuss lifetime of the electron in one layer when there is disorder in other layer.

Using matrix notations defined in Appendix B we have
\begin{equation}
{ \displaystyle {\widetilde{\Sigma}}_1(z) =\displaystyle \sum\limits_{\vec{K_r}} T_+ (\vec{K_r} )\,\,{\mathcal{G}_2} \,(\vec{K_D}+\vec{K_r})\,\,T_- (\vec{K_r} )},
\end{equation}
where $\vec{K_r} $ is the vector of reciprocal lattice which has three values that connect one Dirac point to itself or to the two other equivalent Dirac points.  
$T$ describes the coupling between two layers and the Green operator at wave vector  $\theta\, \vec{\zeta}\times\, \vec{K}_{d \mu}$ is
\begin{equation}
{ \displaystyle {\mathcal{G}_2}\, (z,\theta\, \vec{\zeta}\times\, \vec{K}_{d \mu})=\frac{1}{z-H_-(\theta\, \vec{\zeta}\times\, \vec{K}_{d \mu})}},
\end{equation}
where $\vec{K}_{d \mu}$ counts the three equivalent Dirac points.
And for the Hamiltonian\cite{Omid_these}
\begin{equation}
H_2(\theta\, \vec{\zeta}\times\, \vec{K}_{d \mu}) = 
\left( {\begin{array}{*{20}c}  
  \Delta             &                   -\gamma_0 f\,(\theta\, \vec{\zeta}\times\, \vec{K}_{d \mu})      \\
   - \gamma_0 f^{\ast}\,(\theta\, \vec{\zeta}\times\, \vec{K}_{d \mu})                  &                      \Delta   \\
 \end{array} } \right),
\end{equation}
where $\Delta$ is potential difference between the two layers (layer 1 is in potential 0 and layer 2 is in potential $\Delta$),
$\gamma_0$ is the next-nearest neighbor hoping, and 
\begin{equation}
{\displaystyle f\,(\theta\, \vec{\zeta}\times\, \vec{K_\mu})=2  \sin \frac{\pi \theta}{\sqrt{3}} \,\exp {i \left(\theta_\mu  + \frac{\pi}{2}\,\varepsilon_\theta + \alpha_-(\theta)\right) }},
\end{equation}
%\begin{equation}
%{\displaystyle f\,(\theta\, \vec{\zeta}\times\, \vec{K_\mu})=|f\,(\theta\, \vec{\zeta}\times\, \vec{K_\mu})|\,\,e^{i(\theta_\mu  + %\frac{\pi}{2}\,\varepsilon_\theta \,)  }\,\,e^{i\alpha_-(\theta)}}
%\end{equation}
with $\varepsilon_\theta =\text{sgn}{(\theta)}$  and  $\alpha_- (\theta) ={2 \pi \theta}/{\sqrt{3}} $.
%\begin{equation}
%  \label{nn}
%{\displaystyle |f\,(\theta\, \vec{\zeta}\times\, \vec{K_\mu})|=  \Big{ |}    \gamma_0 e^{ i\frac{2\pi}{\sqrt{3}} \theta } \Big{(} 1- e^{ %-i \frac{2\pi}{\sqrt{3}} \theta } \Big{)}       \Big{|}        = 2  \sin \frac{\pi \theta}{\sqrt{3}} \,\,\,\,\,\,   }
%\end{equation}
Note that this matrix is evaluated at $ \theta\, \vec{\zeta}\times\, \vec{K}_{d \mu} $. Indeed for $ \vec{k} $ sufficiently close to Dirac point $ \vec{k} $, 
because  
%$\hbar  v ( | \vec{k} -\vec{K}_d |) \ll \gamma_0 |f\,(\theta\, \vec{\zeta}\times\, \vec{K}_{d \mu})| $
$\hbar  v ( || \vec{k} -\vec{K}_d ||) \ll \gamma_0 |f\,(\theta\, \vec{\zeta}\times\, \vec{K}_{d \mu})| $
and we can neglect the dependence on the $\vec{k}$ in $H_2$, $\mathcal{G}_2$ and $ \widetilde{\Sigma}_2 (z)$. This corresponds to the general conditions of validity of the present perturbation theory (see above the introduction of appendix \ref{Sec_IC}).

So now after some calculations we get for the self-energy
\begin{equation}
{ \displaystyle {\widetilde{\Sigma}} _0(z)=\sigma(z)\, \bold{I}  },
\end{equation}
%with $ \bold{I}$ is identity matrix and
%\begin{equation}
 %\begin{split}
 %\sigma(z)=\frac{6t^2}{\big{[} z-\gamma_1 g_ (\theta) \big{]}^2 - \gamma_0 ^2 |f(\theta)|^2 }  \\ \times
%  \Big{[}z - \Delta - \gamma_1 g_ (\theta)  -  2 \gamma_0 \sin ( \frac{\pi \theta}{\sqrt{3}} ) \sin (\frac{2\pi \theta}{\sqrt{3}})   \Big{ ]}  
 % \end{split}
%\end{equation}
with
\begin{equation}
{ \displaystyle  \sigma(z)  \simeq \frac{\theta_0 ^2}{  \theta ^2 } \, \Big{[} \Delta - z   \Big{ ]}   },
\label{eq_self}
\end{equation}
where we have introduced $\theta_0 $,
\begin{equation}
%\label{Aformula}
%A= \frac{6t^2}{4/3 \, \pi^2 \gamma_0 ^2}  
\theta_0 = \frac{3}{\sqrt{2}\pi} \frac{t}{\gamma_0}.
\label{equation_t0}
\end{equation}
Using the values of $t \simeq t_1 \simeq 0.12$\,eV (Sec. \ref{Sec_Hamiltonian_R}) and $\gamma_0\simeq 2.7$\, eV one finds that the value of the angle $\theta_0 $  is $\theta_0 \simeq 1.7^\circ$.

%By last equation we can calculate the lifetime of scattering due to disorder (Appendix A)
 
% \begin{equation}
%{ \displaystyle       \frac{1}{\tau_+}= \Im \sigma(z)  \Rightarrow  \frac{1}{\tau_+}  = \frac{6t^2}{ \frac{4\pi^2}{3} \gamma_0 ^2 \theta^2} \frac{\hbar}{\tau_-}    }
%\end{equation}

 % or in a simple form we can write:
  
   %\begin{equation}
%{ \displaystyle       \frac{1}{\tau_+}=\frac{A}{\theta^2} \frac{1}{\tau_-}    }
%\end{equation}

\subsection{Velocity renormalization}
\label{Sec_velocity_renorm}

The eigenvalues are the poles of the Green's function. Therefore the energy $E ( \vec{k} )$ is given by
\begin{equation}
E-\sigma(E)=\pm \hbar v ||\delta \vec{k} ||.
\end{equation}
For $|\vec{k}|=0$, we have solution $E=E_0$ such that
\begin{equation}
 \displaystyle {    E_0-\sigma(E_0)=0  }.
\end{equation}
For small $\vec{k}$, we can write $   E(\vec{k}) = E_0+ \delta E(\vec{k}) $. Eventually we have a nice formula:
\begin{equation}
\delta E = \frac{\pm  \hbar v ||\delta \vec{k} ||}{1-\sigma'(E_0)}.
\end{equation}   
Finally the renormalized velocity $v_r$ is
\begin{equation}
\label{renormalization}
\frac{ v_r }{v} = \frac{  1 }{1+\theta_0^2/ \theta^2 }.     
\end{equation}

\begin{figure}[!h]
\centering
\vspace{1mm}
\scalebox{0.3}{\includegraphics{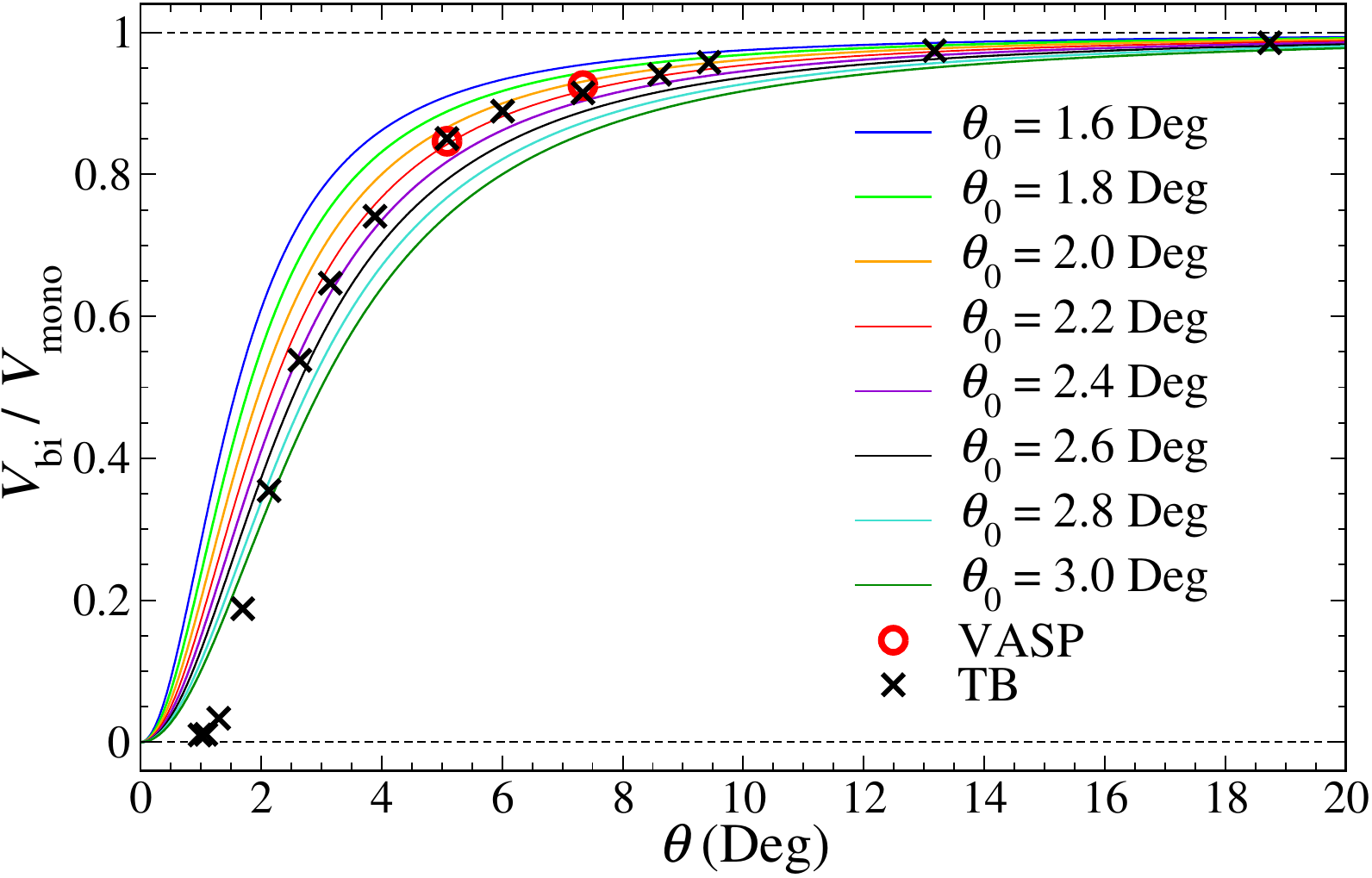}}
\caption{Velocity ratio 
$ v_{r} / v =  V_{\rm bi} / V_{\rm mono} $ 
for commensurate ($n$,$m$) bilayer
cell versus rotation angle $\theta$, computed from equation (\ref{equation_t0}) with varius $\theta_0$ values. 
%The line is the model of Lopez dos Santos et al  
%$ \varv_bi / \varv_{\text{mono}} = 1-9[  \tilde{t}/ (   \varv_{\text{mono}} K \, \sin (\theta/2)    )   ] $, 
%with $ \tilde{t} = 0.11  $eV and
% $ \varv_{\text{mono}} K = 2 \sqrt{3}\gamma_0 \pi=   9.8 $ eV .  
Circle, DFT calculation using VASP code; cross, TB calculations, from Ref. \onlinecite{Trambly10}.      
}
\label{fig:frenormalization}
\end{figure}

Therefore using a well established tight-binding model, we recover velocity renormalization consistent with that of Ref.\,\onlinecite{LopesdosSantos07,LopesdosSantos12}. In addition we find that this velocity renormalization is independent of the difference in potential of two layers.
As it is shown in Fig.\ref{fig:frenormalization},  a systematic study of the renormalization
of the velocity close to the Dirac point is done,\cite{Trambly10} compared to its value in a monolayer graphene, for rotation angles $\theta$ varying
between $0 ^ \circ$ and $60 ^ \circ$ (Fig. \ref{fig:frenormalization}). The renormalization of the
velocity varies symmetrically around $\theta = 30 ^ \circ$. Indeed, the
two limit cases $\theta = 0 ^ \circ$ (AA stacking) and $\theta =60 ^ \circ$ (AB
stacking) are different, but Moir\'{e} patterns when $ \theta \rightarrow   0 ^ \circ$ and
when $ \theta \rightarrow   60 ^ \circ$ are similar because a simple translation by a
vector transforms an AA zone to an AB zone.

Focusing on angles smaller than $ 30 ^ \circ$, 
three regimes can be defined\cite{Trambly10} as a function of the rotation angle $\theta$ (Fig. \ref{fig:frenormalization}). For
large $\theta  (20 ^ \circ \leq \theta \leq 30 ^ \circ   ) $ the Fermi velocity is very close to
that of graphene. 
For intermediate values of $\theta  (3 ^ \circ \leq \theta \leq 20 ^ \circ   )$, the
velocity renormalization is predicted by equation (\ref{renormalization}), as well as by the perturbative theory of Lopes dos Santos et al. \cite{LopesdosSantos12} For the small rotation angles ($\theta < 2^ \circ$) a new regime occurs where  the velocity tends to zero and perturbation theory can not be applied.
 
\subsection{Electron lifetime}
\label{Sec_lifetime}

The two layers of the tBLG can have very different amount of disorder due to their different exposure to environment. For example the lower layer will be in contact with a substrate and the upper layer is exposed either to vacuum or to a gas (sensor application). Therefore it is of high interest to consider the limiting case where defects are present in one layer and absent from the other layer. In the following we consider that defects are present  only in the layer 2. If the two layers were decoupled, defects in one layer would affect electron lifetime in that layer but not in the other one. Since the layers are coupled, defects in one layer will also affect electronic lifetime in the other layer.
In this section, we discuss how such a distribution of defects impacts the electron lifetime. 

If there is disorder in the lower layer (layer 2) the Bloch states of this layer will have a contribution to their self-energy which is imaginary. This can be represented in the simplest possible model by a purely imaginary part of the potential energy $\Delta$,
\begin{equation}
\displaystyle{  \Delta = -\frac{i\hbar}{\tau_2}  } \, ,
\end{equation}   
where $ \tau_2 $ is the lifetime in the layer 2 due to disorder in the layer 2. Using formula (\ref{eq_self}), we see that electrons in the layer 1 acquire an imaginary self-energy
\begin{equation}
\displaystyle{ \Im \sigma(z)= - \frac{i \hbar}{\tau_1}  = - \frac{i \hbar}{\tau_2}  \frac{\theta_0 ^2}{  \theta ^2 }   } .
\end{equation}   
Therefore the lifetimes $\tau_1$ and $\tau_2$ in the layer 1 and layer 2 are related through: 
\begin{equation}
\displaystyle{          \frac{\tau_1}{\tau_2} =  \left( \frac{\theta}{\theta_0} \right)^2 } \, ,
\label{eq_rapport_tau}
\end{equation}   
where $\theta_0$  is given by equation (\ref{equation_t0}), and is same quantity as in the velocity renormalization expression (\ref{renormalization}).

\subsection{Spatial  variation of density of states}
\label{Sec_SpatialDOS}

As explained above the coupling between Bloch states of different wave-vectors in layer 1 (due to interlayer coupling with layer 2) corresponds to the selection rule
\begin{equation}
\delta \vec{k_f} \simeq \delta \vec{k_0} +\vec{G_k} ,
\label{SRdKG11}
\end{equation}
where $\vec{G_k}=\theta \vec{\zeta} \times (\vec{K}_{Di} -\vec{K}_{Dj})$ is a reciprocal lattice vector of the Moir\'{e} lattice.
The typical difference in energy  between  Bloch states of  $\vec {\delta k_f}$ and of $\vec {\delta k_0}$ is 
$\Delta E \simeq \hbar v \theta ||\vec{K}_{Di}||$.
 This difference is nearly independent of $\vec{\delta k_0}$ provided that it is sufficiently close to zero. The typical coupling  is $t_{eff}\simeq t^2/(\hbar v \theta ||\vec{K}_{Di}||)$. 

Then the mixing between states of wave vector close to ($\vec{K}_{Di}$) and wave vector close to $\vec{K}_{Di}+\vec{G_k} $ will be of order $t_{eff}/ \Delta E$ i.e. of order $(\theta_1/\theta)^2$. 
%We note that the mixing is independent of $\delta\vec{ k_0}$ provided that it is sufficiently close to zero. 
Therefore the relative variation of the DOS of a state is independent of the energy, for states sufficiently close to the Dirac point, and it depends only on the position in the Moir\'{e} pattern. A precise calculation\cite{Omid_these} provides the expression given in the main text (equation (\ref{eqLDOS})),
\begin{equation}
\frac{\Delta \rho(E,\vec{r} )}{\rho_m(E)} \simeq  \left( \frac{\theta_{1}}{\theta} \right)^2  \sum\limits_{j=1}^{6}  \, \cos(\vec{G_j} \cdot \vec{r}),
%\frac{\Delta \rho(E,\vec{r} )}{\rho(E)} \simeq %\cong  
%\left( \frac{\theta_{1}}{\theta} \right)^2  \sum\limits_{j=1}^{6}  \,   
%\cos(\vec{G_j} \cdot \vec{r}),
\label{eqLDOS2}
\end{equation}
where $\vec{G_j}$ are 6 equivalent vectors of the reciprocal space of the Moir\'{e} lattice and where the  rotation angle $\theta_{1}$ is given by  
 \begin{equation}
\theta_{1}=\frac{\sqrt{2}t}{(\hbar v K_D )} = \frac{\theta_0}{\sqrt{3}}.
\label{apptheta}
\end{equation}
Using the interlayer coupling value $t \simeq 0.12$\,eV (Appendix Sec. \ref{Sec_Hamiltonian_R}) one finds that  $\theta_{1}$ is close to one degree.

We emphasize that the present theory is perturbative in the coupling $t$. This perturbation theory is valid for sufficiently large values of $\theta$ as explained in the introduction of appendix (\ref{Sec_IC}). The other assumption is to neglect Fourier components of the interlayer Hamiltonian that couple a Bloch state to other states having wave vectors away from the Dirac cones. This approximation can lead to the under estimation of modulations of the DOS at spatial frequencies high with respect to the Moir\'{e} period. This could explain why 
the DOS modulation (TB calculations) on sub-lattices A and B can differ by about $\pm 15 \%$ as compared to averaged DOS, whereas the present perturbative theory does not predict this difference. Note that the average DOS of two neighboring A and B atoms is well reproduced by the analytical model. 

%Indeed in the calculation by Castro Neto et. al \cite{LopesdosSantos12} the authors retain also only one Fourier component of the transfer matrix $t(\vec{k})$, but do numerical non-perturbative calculation. They find also a DOS which is the same for neighboring A and B atoms.

\bibliography{biblio}

\end{document}